\documentclass[usenatbib]{mn2e}
\usepackage{graphicx}

\title[Photometric study of new southern SU UMa-type dwarf novae]
{Photometric study of new southern SU UMa-type dwarf novae and candidates:
V877 Ara, KK Tel and PU CMa}

\author[T. Kato et al.]
{\parbox[t]{\textwidth}{
       Taichi Kato$^1$,
       Roland Santallo$^2$,
       Greg Bolt$^3$,
       Tom Richards$^4$,
       Peter Nelson$^5$, \\
       Berto Monard$^6$, 
       Makoto Uemura$^1$,
       Seiichiro Kiyota$^7$,
       Rod Stubbings$^8$, \\
       Andrew Pearce$^9$,
       Tsutomu Watanabe$^{10}$, 
       Patrick Schmeer$^{11}$,
       Hitoshi Yamaoka$^{12}$ \\
} \\
       $^1$ Department of Astronomy, Faculty of Science,
       Kyoto University, Sakyo-ku, Kyoto 606-8502 Japan \\
       $^2$ Southern Stars Observatory, Po Box 60972, 98702 FAAA TAHITI,
       French Polynesia \\
       $^3$ 295 Camberwarra Drive, Craigie, Western Australia 6025,
       Australia \\
       $^4$ Woodridge Observatory, 8 Diosma Rd, Eltham, Vic 3095, Australia \\
       $^5$ RMB 2493, Ellinbank 3820, Australia \\
       $^6$ Bronberg Observatory, PO Box 11426, Tiegerpoort 0056,
       South Africa \\
       $^7$ Variable Star Observers League in Japan (VSOLJ),
       1-401-810 Azuma, Tsukuba, 305-0031, Japan \\
       $^8$ 19 Greenland Drive, Drouin 3818, Victoria, Australia \\
       $^9$ 32 Monash Ave, Nedlands, WA 6009, Australia \\
       $^{10}$ Variable Star Observers League in Japan (VSOLJ),
       117 Shirao dormitory, 1414 Oonakazato, Shizuoka 418-0044, Japan \\
       $^{11}$ Bischmisheim, Am Probstbaum 10, 66132 Saarbr\"{u}cken,
       Germany \\
       $^{12}$ Faculty of Science, Kyushu University, Fukuoka 810-8560,
       Japan
}

\date{Accepted.
      Received;
      in original form}

\pagerange{\pageref{firstpage}--\pageref{lastpage}}
\pubyear{2002}

\begin{document}

\maketitle

\label{firstpage}

\begin{abstract}
   We photometrically observed three dwarf novae V877 Ara, KK Tel and
PU CMa.  We discovered undisputed presence of superhumps in V877 Ara and
KK Tel, with mean periods of 0.08411(2) d and 0.08808(3) d, respectively.
Both V877 Ara and KK Tel are confirmed to belong to long-period SU UMa-type
dwarf novae.  In V877 Ara, we found a large decrease of the superhump
period ($\dot{P}/P$ = $-$14.5$\pm$2.1 $\times$ 10$^{-5}$).  There is
evidence that the period of KK Tel decreased at a similar or a more
exceptional rate.  Coupled with the past studies of superhump period
changes, these findings suggest that a previously neglected diversity of
phenomena is present in long-period SU UMa-type dwarf novae.
The present discovery of a diversity in long-period SU UMa-type systems
would become an additional step toward a full understanding the dwarf nova
phenomenon.  PU CMa is shown to be an excellent candidate for an SU UMa-type
dwarf nova.  We examined the outburst properties of these dwarf novae,
and derived characteristic outburst recurrence times.  Combined with the
recently published measurement of the orbital period of PU CMa,
we propose that PU CMa is the first object filling the gap between the
extreme WZ Sge-type and ER UMa-type stars.
\end{abstract}


\section{Introduction}

   Dwarf novae are a class of cataclysmic variables (CVs), which are
close binary systems consisting of a white dwarf and a red dwarf secondary
transferring matter via the Roche-lobe overflow (for recent reviews,
see \cite{war95book,hel01book}).  There exists a class
of dwarf novae, called SU UMa-type dwarf novae.  All SU UMa-type dwarf
novae show superhumps during their long, bright outbursts (superoutbursts).
[For a recent review of dwarf novae and SU UMa-type dwarf novae,
see \citet{osa96review} and \citet{war95suuma}, respectively.]
Superhumps are 0.1--0.5 mag modulations which have periods
(superhump period: $P_{\rm SH}$) a few percent
longer than the system orbital period ($P_{\rm orb}$).  The difference
between $P_{\rm SH}$ and $P_{\rm orb}$ is understood as a consequence
of the apsidal motion \citep{osa85SHexcess,mol92SHexcess}
of a tidally induced eccentric accretion disk
\citep{whi88tidal,hir90SHexcess,lub91SHa}.  The origin of superhumps
is explained as increased viscous dissipation around periodic conjunctions
between the major axis of the elongated accretion disk and the
secondary star.  This explanation has been recently confirmed with
more detailed hydrodynamical simulations
\citep{mur96SPHtidal,mur98SH,mur00SHprecession}.
\citet{tru00DNoutburst,tru01DNsuperoutburst} indeed succeeded in reproducing
the light curves of dwarf novae with hydrodynamical simulations
within the thermal-tidal disk instability model
\citep{osa89suuma,osa96review}.

   Early development in study of SU UMa-type dwarf novae largely owed
to southern bright objects (VW Hyi, Z Cha, OY Car etc. see e.g.
\cite{vog74vwhyi,vog80suumastars,vog81oycar,vog82zcha,vog83oycar,
vog83lateSH,war74vwhyi,war74zcha,war75v436cen,war85suuma,coo81zchapdot,
coo84zcha,odo86zchadiskradius,sma79zcha,sma85SHzcha,bai81oycar}) with
dedicated telescopes.
The recent advent of the wide availability of CCDs, however, has opened
a new window to study of SU UMa-type dwarf novae.  Excellent examples
include (a) an extension to faint objects (e.g.
\cite{how88faintCV1,how90faintCV3,how91faintCV4,muk90faintCV,szk89faintCV2})
and (b) timely observations of SU UMa-type outbursts.
These new techniques have produced a number of striking discoveries
or new concepts: e.g. recognition of a class of short-period, rarely
outbursting SU UMa-type dwarf novae \citep{how95TOAD}, establishment
the concept of WZ Sge-type dwarf novae
\citep{kat96alcom,kat97egcnc,kat01hvvir,how96alcom,pat96alcom,pat98egcnc,
nog97alcom,mat98egcnc,ish01rzleo}, discovery of peculiar
ER UMa-type stars \citep{kat95eruma,rob95eruma,nog95v1159ori,
nog95rzlmi,pat95v1159ori,kat96diuma}, and discovery of ultra-short period
SU UMa-type dwarf novae breaking the standard evolutionary sequence of
short-period CVs \citep{uem02j2329letter,ski02j2329}.  These discoveries
have dramatically changed our understanding of CVs.
In recent years, \citet{wou01v359cenxzeriyytel} applied a technique of
time-resolved CCD photometry, which was largely introduced by
\citet{how88faintCV1,szk89faintCV2}, to the previously unexploited field
of faint southern CVs, and has indeed proven the high productivity of
such a study.

   In the most recent years, the advent of the Internet, wide availability
of CCDs and a global network of observers, best exemplified by
the VSNET Collaboration\footnote{
http://www.kusastro.kyoto-u.ac.jp/vsnet/.} has enabled a rapid circulation
of outburst alerts and opened a new window to globally and timely
study the novel field of {\it transient object astronomy}.  Combined with
the established technique of time-resolved CCD photometry of outbursting
dwarf novae, this kind of global collaboration has become a breakthrough
in study of CVs (a wealth of scientific highlights is presented
on the VSNET webpage).  In this paper, we present the very first result
of this global network on newly discovered outbursts of southern CVs.

\section{CCD Observation}

   The observers, equipment and reduction software are summarized in
Table \ref{tab:equipment}.  The Kyoto observations were analyzed using the
Java$^{\rm TM}$-based PSF photometry package developed by one of the
authors (TK).  The other observers performed aperture photometry.
The observations used unfiltered CCD systems having a response
close to Kron-Cousins $R_{\rm c}$ band for outbursting dwarf novae,
except for the $V$-band
observation by SK.  The errors of single measurements are typically
less than 0.01--0.03 mag unless otherwise specified.

\begin{figure*}
  \includegraphics[angle=0,width=16cm]{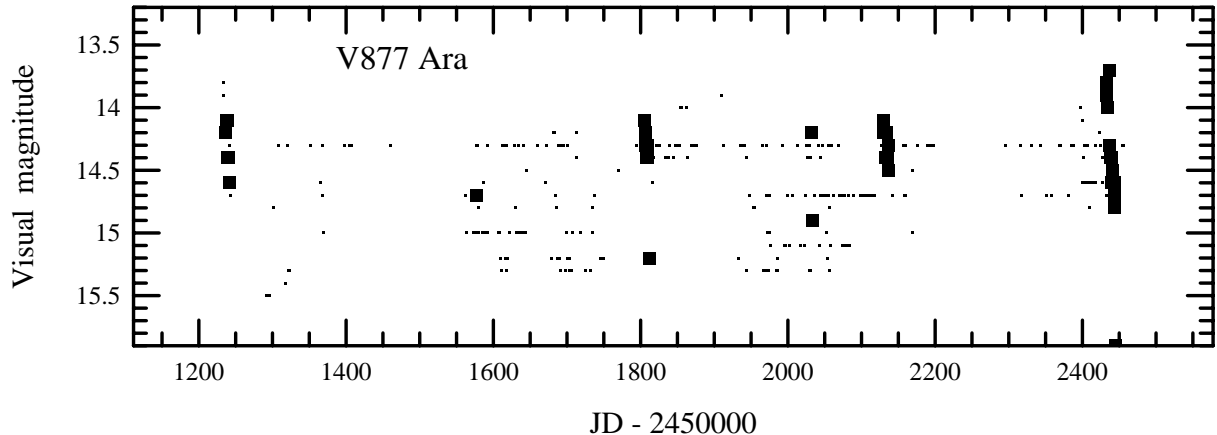}
  \caption{Long-term visual light curve of V877 Ara.  Large and small dot
  represent positive and negative (upper limit) observations, respectively.
  Four major outbursts occurring at JD 2451236, 2451805, 2452129 and
  2452433 are superoutbursts.}
  \label{fig:v877long}
\end{figure*}

\begin{table}
\caption{Observers and Equipment.} \label{tab:equipment}
\begin{center}
\begin{tabular}{cccc}
\hline\hline
Observer   & Telescope &  CCD  & Software \\
\hline
Santallo   & 20-cm SCT & ST-7E & AIP4Win \\
Bolt       & 25-cm SCT & ST-7  & MuniPack$^a$ \\
Richards   & 18-cm refractor & ST-7E & AIP4Win$^b$ \\
Nelson     & 32-cm reflector & ST-8E & AIP4Win \\
Monard     & 30-cm SCT & ST-7E & AIP4Win \\
Kyoto      & 25-cm SCT & ST-7  & Java$^c$ \\
Kiyota     & 25-cm SCT & AP-7 + $V$ filter & MIRA A/P \\
\hline
 \multicolumn{4}{l}{$^a$ http://munipack.astronomy.cz.} \\
 \multicolumn{4}{l}{$^b$ MaxIm/DL was used for KK Tel.} \\
 \multicolumn{4}{l}{$^c$ See text.} \\
\end{tabular}
\end{center}
\end{table}

   Barycentric corrections to the observed times were applied before the
following analysis.

\section{V877 A\lowercase{ra} = NSV 08383}

\subsection{Introduction}

   V877 Ara was originally discovered as a suspected variable star
(CSV 7612 = NSV 08383).  The star was selected as a possible dwarf nova in
\citet{vog82atlas}.  From the finding chart presented in this literature,
\citet{lop85CVastrometry} provided astrometry of the proposed quiescent
counterpart at 17$^h$ 16$^m$ 58$^s$.96, $-$65$^{\circ}$ 33$'$ 00$''$.2
(precessed to J2000.0).  The star has been intensively monitored by members
of the VSNET Collaboration.  The first outburst was detected by B. Monard
on 1999 February 27 at visual magnitude 14.2 (vsnet-alert 2715)\footnote{
http://www.kusastro.kyoto-u.ac.jp/vsnet/Mail/alert2000/\\msg00715.html.
}.  Monard further noticed that the outbursting object is different from
the proposed quiescent counterpart (vsnet-alert 2727)\footnote{
http://www.kusastro.kyoto-u.ac.jp/vsnet/Mail/alert2000/\\msg00727.html.
}.  Although no detailed time-resolved photometry was performed, the
overall behavior of the 1999 February--March outburst resemble that of
a superoutburst of an SU UMa-type dwarf nova (vsnet-alert 4110)\footnote{
http://www.kusastro.kyoto-u.ac.jp/vsnet/Mail/alert4000/\\msg00110.html.
}.

\subsection{Long-term Light Curve}\label{sec:v877long}

   Figure \ref{fig:v877long} shows the long-term visual light curve of
V877 Ara.  Four major outbursts occurring at JD 2451236, 2451805, 2452129
and 2452433 are clearly seen.  As shown in section \ref{sec:v877SH},
the major outburst around JD 2452433 is a superoutburst as demonstrated
by the secure detection of superhumps, thereby establishing the SU UMa-type
nature of V877 Ara.  The earlier three outbursts had durations longer
than 5 d (section \ref{sec:v877suuma}), and comparable
maximum magnitudes to that of the JD 2452433 outburst.  Since all
well-observed long (more than 5 d) outbursts of SU UMa-type dwarf novae
have been confirmed to be superoutbursts \citep{vog80suumastars,war85suuma},
we can safely conclude that these outbursts are indeed superoutbursts.
Two shorter outbursts were detected between these outbursts.  These
outbursts are most likely normal outbursts.  The overall characteristics
of the long-term light curve is that of a typical SU UMa-type dwarf nova.

\subsection{The 2002 June Superoutburst}

   The 2002 June outburst was detected by R. Stubbings on June 7 at
visual magnitude 13.8.  The object was reported to be below the detection
limit on the previous night; the outburst must have been caught during
its earliest stage.  We undertook time-resolved CCD photometry
campaign.  The magnitudes were measured relative to
GSC 9060.112, whose constancy during the observation was confirmed
by a comparison with GSC 9060.610.  The log of observations is 
given in Table \ref{tab:v877log}.

\begin{table}
\caption{Journal of the 2002 CCD photometry of V877 Ara.}\label{tab:v877log}
\begin{center}
\begin{tabular}{crccrc}
\hline\hline
\multicolumn{2}{c}{2002 Date}& Start--End$^a$ & Exp(s) & $N$
        & Obs$^b$ \\
\hline
June &  9 & 52434.951--52435.110 &  50  & 194 & S \\
     & 11 & 52436.959--52437.123 &  50  & 163 & S \\
     & 11 & 52437.069--52437.283 &  60  & 273 & B \\
     & 15 & 52441.010--52441.219 &  75  & 215 & B \\
     & 17 & 52442.920--52443.222 & 240  &  93 & R \\
     & 17 & 52443.105--52443.285 & 120  & 120 & B \\
\hline
 \multicolumn{6}{l}{$^a$ BJD$-$2400000.} \\
 \multicolumn{6}{l}{$^b$ S (Santallo), B (Bolt), R (Richards)} \\
\end{tabular}
\end{center}
\end{table}

\begin{figure}
  \includegraphics[angle=0,width=8.8cm]{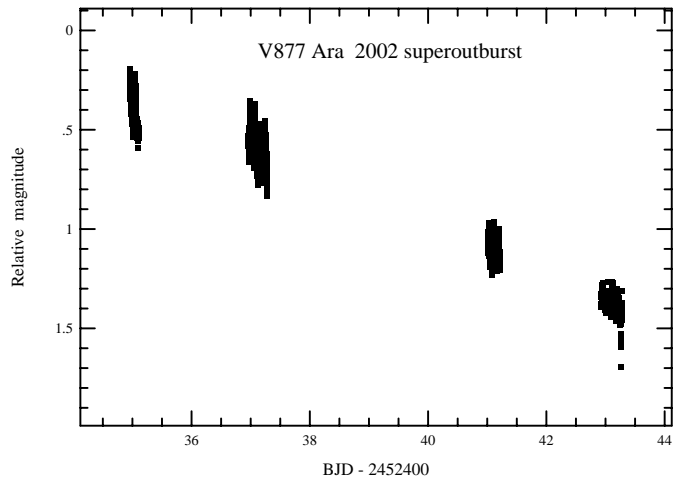}
  \caption{Light curve of the 2002 June superoutburst of V877 Ara.
  The magnitudes are given relative to GSC 9060.112 and are on a system
  close to $R_{\rm c}$.}
  \label{fig:v877lc}
\end{figure}

\begin{figure}
  \includegraphics[angle=0,width=8.8cm,height=11cm]{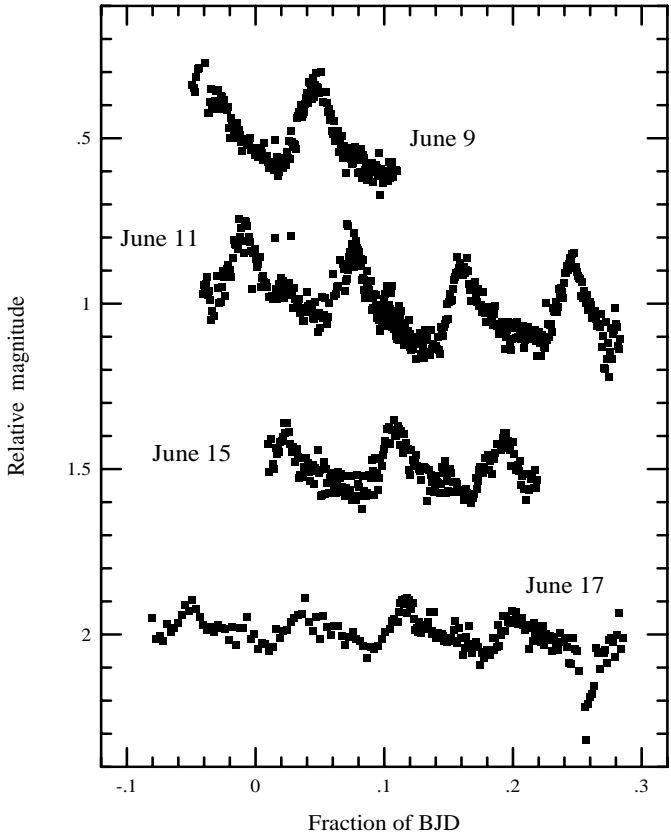}
  \caption{Nightly light curves of V877 Ara.  Gradually decaying
  superhumps are clearly visible on all nights.}
  \label{fig:v877night}
\end{figure}

   Figures \ref{fig:v877lc} and \ref{fig:v877night} are the overall
and nightly light curves of V877 Ara.  Gradually decaying superhumps
are clearly visible on all nights; this confirms the SU UMa-type
nature of V877 Ara.  The overall light curve shows a linear decline
with a rate of 0.12 mag d$^{-1}$.  This value is quite characteristic
of a normal SU UMa-type dwarf nova.  The dip-like structure
near the end of the June 17 run may have been a result of slightly
unfavorable observing condition.

\subsection{Superhump Period}\label{sec:v877SH}

\begin{figure}
  \includegraphics[angle=0,width=8cm]{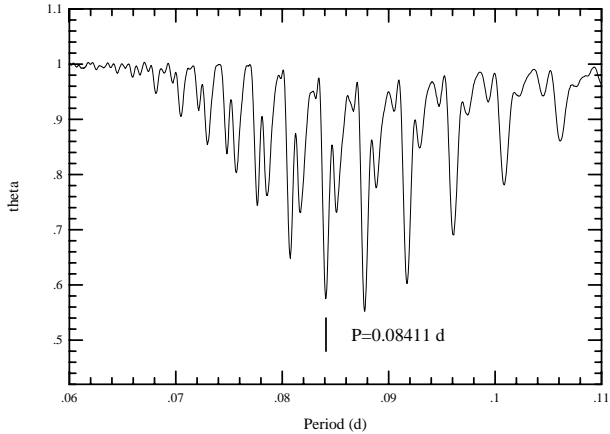}
  \caption{Period analysis of V877 Ara.  See text for the explanation of
  the selection of the correct alias ($P_{\rm SH}$ = 0.08411 d).}
  \label{fig:v877pdm}
\end{figure}

\begin{figure}
  \includegraphics[angle=0,width=8cm]{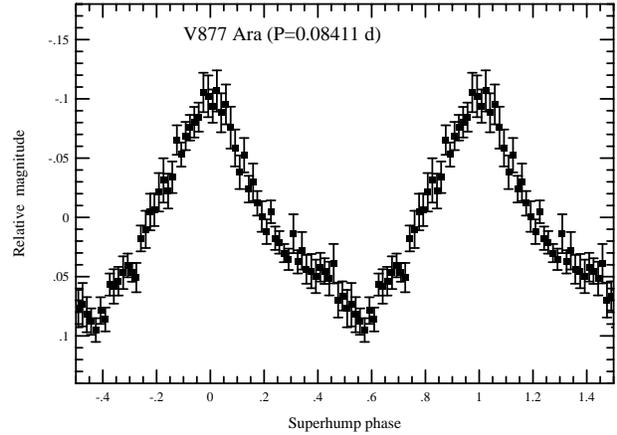}
  \caption{Mean superhump profile of V877 Ara.}
  \label{fig:v877ph}
\end{figure}

   Figure \ref{fig:v877pdm} shows the result of a period analysis using
Phase Dispersion Minimization (PDM: \cite{PDM}) applied to the entire
data set after removing the linear decline trend.  The selection
of the correct period among possible aliases was performed using an
independent analysis of the longest run (June 11), which yielded a period
of 0.0853(4) d.\footnote{The quoted error from a single-night observation
  should be regarded as an approximation rather than the true error, since
  an estimated of such a period may have been affected by an uncorrected
  systematic variation of the observing condition or an intrinsic variation
  in the waveform.  We therefore regard that the seeming difference between
  nightly-based periods on June 10 and June 11 is not significant.
  This error, however, safely rejected the longer possible alias by at least
  6$\sigma$.
}
This selection was also confirmed by a comparison with
the independently obtained period (0.0842 d) communicated from S. Walker
based on his June 10 observation.  The adopted mean period during the
superoutburst is 0.08411(2) d.  A relatively large difference between the
mean period and a period on a single-night basis may be ascribed to
the large superhump period change described later.
Figure \ref{fig:v877ph} shows the
phase-averaged profile of superhumps.  The rapidly rising and slowly
fading superhump profile is characteristic of an SU UMa-type
dwarf nova \citep{vog80suumastars,war85suuma}.

   We extracted the maxima times of superhumps from the light curve by eye.
The averaged times of a few to several points close to the maximum were
used as representatives of the maxima times.  The errors of the maxima
times are usually less than $\sim$0.002 d.  The resultant superhump maxima
are given in Table \ref{tab:v877max}.  The values are given to 0.0001 d in
order to avoid the loss of significant digits in a later analysis.
The cycle count ($E$) is defined as the cycle number since BJD 2452434.961.
A linear regression to the observed superhump times gives the following
ephemeris:

\begin{equation}
{\rm BJD (maximum)} = 2452434.9715(31) + 0.084025(47) E. \label{equ:reg1}
\end{equation}

\begin{figure}
  \includegraphics[angle=0,width=8cm]{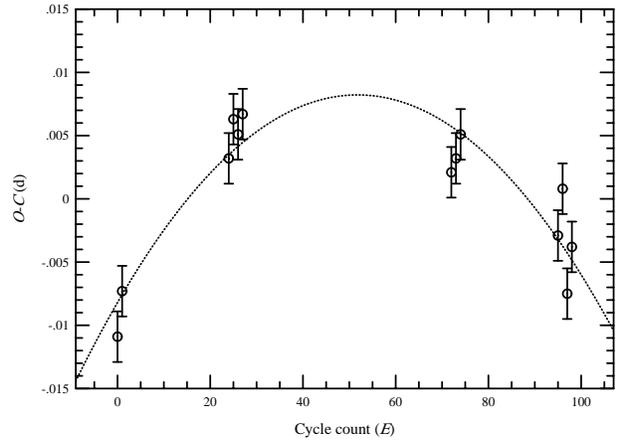}
  \caption{$O-C$ diagram of superhump maxima of V877 Ara.  The error
  bars correspond to the upper limits of the errors.
  The parabolic fit corresponds to equation \ref{equ:reg2}}
  \label{fig:v877oc}
\end{figure}

   Figure \ref{fig:v877oc} shows the ($O-C$)'s against the mean superhump
period (0.084025 d) from a linear regression (equation \ref{equ:reg1}).
The diagram clearly shows the decrease in the superhump period throughout
the superoutburst plateau.  The times of the superhump maxima in this interval
can be well represented by the following quadratic equation:

\begin{eqnarray}
{\rm BJD} & {\rm (maximum)} = 2452434.9633(18) + 0.084659(92) E \nonumber \\
    & -6.12(87) \times 10^{-6} E^2. \label{equ:reg2}
\end{eqnarray}

   The quadratic term corresponds to $\dot{P}$ = $-$12.2$\pm$1.7 $\times$
10$^{-6}$ d cycle$^{-1}$, or $\dot{P}/P$ = $-$14.5$\pm$2.1 $\times$ 10$^{-5}$.
This measured quadratic term is one of the extremely negative values
among all SU UMa-type dwarf novae \citep{kat01hvvir}.

\begin{table}
\caption{Times of superhump maxima of V877 Ara.}\label{tab:v877max}
\begin{center}
\begin{tabular}{ccc}
\hline\hline
$E^a$  & BJD$-$2400000 & $O-C^b$ \\
\hline
  0 & 52434.9606 & $-$0.0109 \\
  1 & 52435.0482 & $-$0.0073 \\
 24 & 52436.9913 &  0.0032 \\
 25 & 52437.0784 &  0.0063 \\
 26 & 52437.1613 &  0.0051 \\
 27 & 52437.2469 &  0.0067 \\
 72 & 52441.0234 &  0.0021 \\
 73 & 52441.1086 &  0.0032 \\
 74 & 52441.1945 &  0.0051 \\
 95 & 52442.9510 & $-$0.0029 \\
 96 & 52443.0387 &  0.0008 \\
 97 & 52443.1145 & $-$0.0075 \\
 98 & 52443.2022 & $-$0.0038 \\
\hline
 \multicolumn{3}{l}{$^a$ Cycle count since BJD 2452434.961.} \\
 \multicolumn{3}{l}{$^b$ $O-C$ calculated against equation
                    \ref{equ:reg1}.} \\
\end{tabular}
\end{center}
\end{table}

\subsection{Astrometry and Quiescent Counterpart}

\begin{figure}
  \begin{center}
  \includegraphics[angle=0,width=8cm]{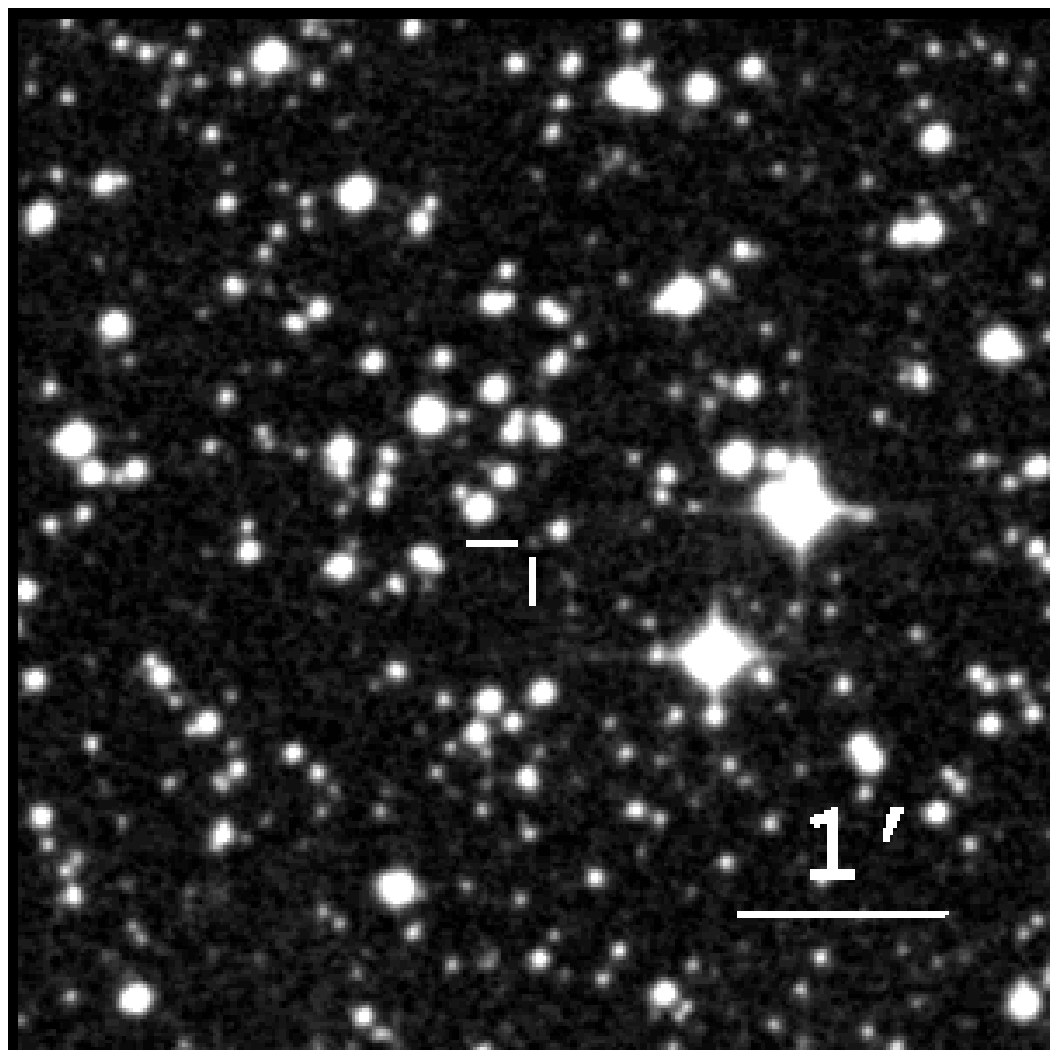} \\
  \includegraphics[angle=0,width=8cm]{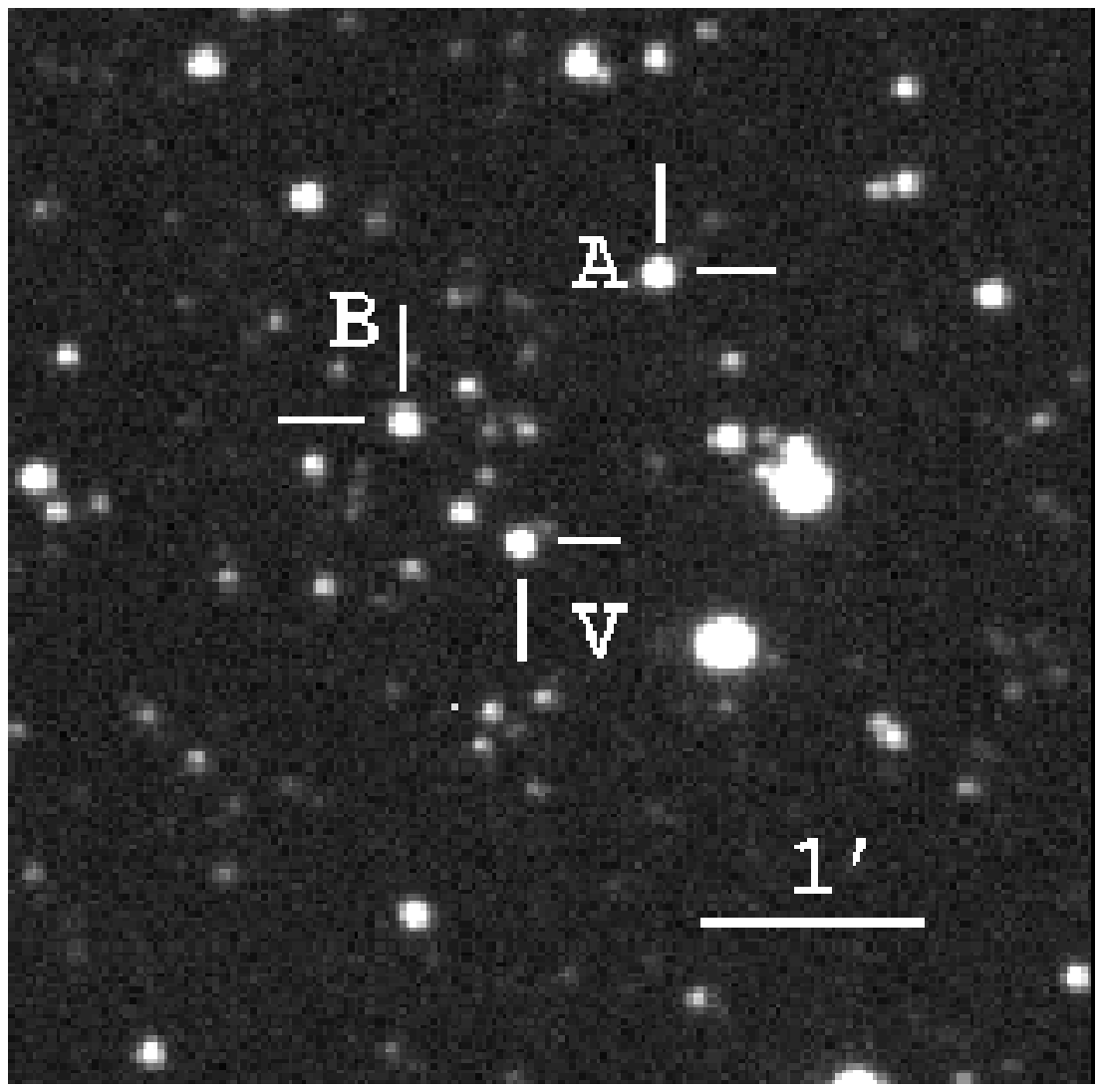}
  \end{center}
  \caption{Identification of V877 Ara.  (Upper) In quiescence,
  reproduced from the DSS 2 red image.  (Lower) In outburst, taken on
  2002 June 09 by BM.  V = V877 Ara, A = GSC 9060.112, B = GSC
  9060.610 (see text for the details).}
  \label{fig:v877id}
\end{figure}

   Astrometry of the outbursting V877 Ara was performed on CCD
images taken by R. Santallo and B. Monard.  An average of measurements
of seven images (UCAC1 system, 60 -- 150 reference stars; internal
dispersion of the measurements was $\sim$0$''$.03) has yielded
a position of 17$^h$ 16$^m$ 53$^s$.936, $-$65$^{\circ}$ 32$'$ 51$''$.49
(J2000.0).  The position agrees with the USNO$-$A2.0 star at
17$^h$ 16$^m$ 53$^s$.926, $-$65$^{\circ}$ 32$'$ 51$''$.80
(epoch 1979.891 and magnitudes $r$ = 17.8, $b$ = 19.1), which is most
likely the quiescent counterpart of V877 Ara (Figure \ref{fig:v877id}).
Comparing with the USNO$-$A2.0 position and the DSS 1 images (epoch =
1975 -- 1978), no apparent proper motion was detected; its upper
limit is deduced to be 0$''$.04 yr$^{-1}$.
No counterpart is recorded in GSC$-$2.2, UCAC and 2MASS catalogs.

\subsection{V877 Ara as an SU UMa-type Dwarf Nova}\label{sec:v877suuma}

   Table \ref{tab:v877out} lists the recorded outbursts of V877 Ara.
The durations of outbursts have a clear bimodal ($\leq$2 d or $\geq$5 d)
distribution, which is very characteristic of an SU UMa-type star
\citep{vog80suumastars,war85suuma}.  The observed intervals
between superoutbursts implies the existence of a missed superoutburst
around the solar conjunction between JD 2451236 and JD 2451805.
The intervals between superoutbursts (supercycle length) are thus 285--374 d.
The observed number of normal outbursts between superoutbursts does not
seem to be exceptionally low for an SU UMa-type star
\citep{war95suuma,nog97sxlmi} considering the faintness (14.2--14.6 mag)
of normal outbursts, which is close to the detection limit.  Future
deep monitoring is encouraged to determine the true cycle length
of normal outbursts.

\begin{table}
\caption{List of Outbursts of V877 Ara.}\label{tab:v877out}
\begin{center}
\begin{tabular}{ccccc}
\hline\hline
JD start$^a$ & JD end$^a$ & Max & Duration (d) & Type \\
\hline
51236.6 & 51241.4 & 14.1 & $>$5 & super \\
51577.3 &   --    & 14.6 & 1    & normal \\
51805.2 & 51811.9 & 14.1 & $>$7 & super \\
52032.3 & 52033.3 & 14.2 & 2    & normal \\
52129.2 & 52137.2 & 14.1 & $>$8 & super \\
52432.9 & 52445.3 & 13.9 & 13   & super \\
\hline
 \multicolumn{5}{l}{$^a$ JD$-$2400000.} \\
\end{tabular}
\end{center}
\end{table}

\section{KK T\lowercase{el}}

\subsection{Introduction}

   KK Tel is a dwarf nova discovered by C. Hoffmeister.  \citet{vog82atlas}
provided a finding chart.  \citet{how90highgalCV} suggested that KK Tel
may be a member of high galactic latitude CVs, which were then considered
to be a new population of CVs with large outburst amplitudes
\citep{how95TOAD}.  \citet{how91faintCV4} obtained CCD time-resolved
photometry of KK Tel in quiescence, and recorded regularly recurring humps
with a period of 0.084 d.  Among the CVs studied by \citet{how91faintCV4},
KK Tel showed most prominent humps, which resembled those of
a high-inclination eclipsing CV, DV UMa
\citep{how87dvuma,how88dvuma,how93arcncaypscdvuma}.  The characteristics
of the humps in KK Tel was further discussed in \citet{how96tvcrv}.
\citet{zwi95CVspec2} obtained a spectrum in quiescence, and confirmed
the presence of Balmer emission lines.  This observation spectroscopically
confirmed the CV nature of KK Tel.  Since the reported period strongly
suggests the SU UMa-type nature, we undertook a photometric campaign.

\subsection{The 2002 June Superoutburst}

   The 2002 June outburst was detected by R. Stubbings on June 17 at
visual magnitude 13.8.  Due to a 4-d gap of observation before this
detection, the exact epoch of the start of the outburst was not
determined.  We started a CCD photometric campaign following this
detection.  Each observer measured relative magnitudes against a constant
star in the same field.  Since different observers used different
comparison stars and the accurate adjustment of the zero point is
difficult because of
the presence of a close visual companion to KK Tel, we simply subtracted
a nightly average from each runs and the combined data were subject to period
analyses.  The log of observations is given in Table
\ref{tab:kklog}.

\begin{table}
\caption{Journal of the 2002 CCD photometry of KK Tel.}\label{tab:kklog}
\begin{center}
\begin{tabular}{crccrc}
\hline\hline
\multicolumn{2}{c}{2002 Date}& Start--End$^a$ & Exp(s) & $N$
        & Obs$^b$ \\
\hline
June & 18 & 52443.969--52444.070 &  45 & 133 & N \\
     & 20 & 52445.946--52446.228 & 120 & 171 & R \\
     & 22 & 52448.075--52448.151 &  30 & 136 & S \\
\hline
 \multicolumn{6}{l}{$^a$ BJD$-$2400000.} \\
 \multicolumn{6}{l}{$^b$ N (Nelson), R (Richards), S (Santallo)} \\
\end{tabular}
\end{center}
\end{table}

\subsection{Superhump Period}

\begin{figure}
  \includegraphics[angle=0,width=8cm,height=10cm]{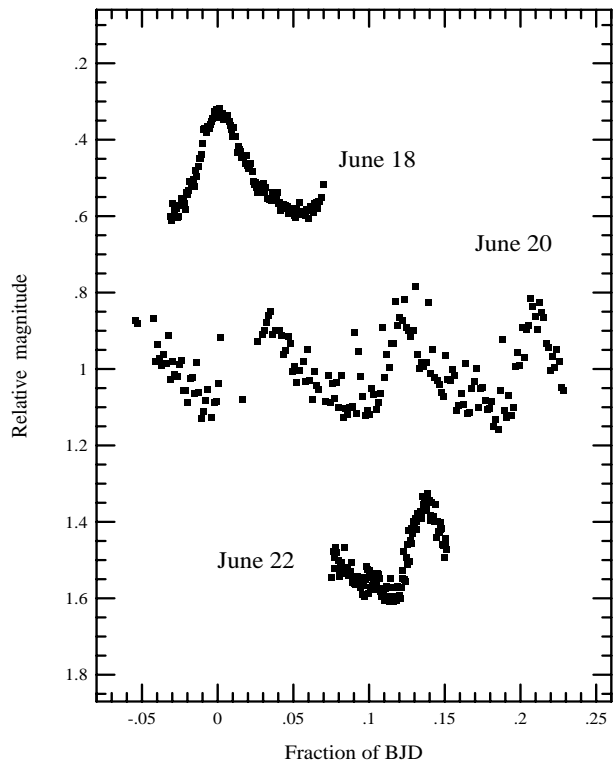}
  \caption{Nightly light curves of KK Tel.  Superhumps are clearly
  visible on all nights.}
  \label{fig:kknight}
\end{figure}

\begin{figure}
  \includegraphics[angle=0,width=8cm]{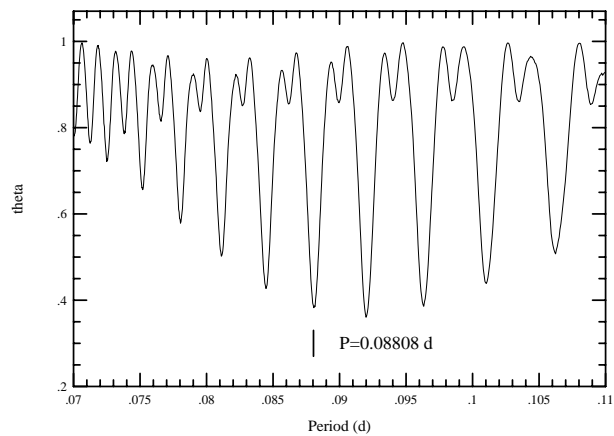}
  \caption{Period analysis of KK Tel.  See text for the explanation of
  the selection of the correct alias ($P_{\rm SH}$ = 0.08808 d).}
  \label{fig:kkpdm}
\end{figure}

   Figure \ref{fig:kknight} shows nightly light curves of KK Tel.
Superhumps are clearly visible on all nights.  This observation has finally
established the SU UMa-type nature of KK Tel.
The relatively large scatter on June 20 may have been caused by an
interference by the nearby companion.
Figure \ref{fig:kkpdm} shows the result of PDM period analysis.
The selection of the correct period among possible aliases was performed
using an independent analysis of the longest run (June 20), which yielded
a period of 0.0878(10) d.  The identification of the superhump period
is also confirmed by a comparison with the quiescent photometric period
of 0.084 d \citep{how91faintCV4}.  No other alias gives a reasonable match
with this quiescent period.  The best determined mean $P_{\rm SH}$ is
0.08808(3) d.

   We extracted the maxima times of superhumps from the light curve
just as in section \ref{sec:v877SH}.  The resultant superhump maxima
are given in Table \ref{tab:kkmax}.  A linear regression of the observed
times is given in equation \ref{equ:kkreg1}.

\begin{table}
\caption{Times of superhump maxima of KK Tel.}\label{tab:kkmax}
\begin{center}
\begin{tabular}{ccc}
\hline\hline
$E^a$  & BJD$-$2400000 & $O-C^b$ \\
\hline
  0 & 52444.0018 & $-$0.0055 \\
 23 & 52446.0351 &  0.0036 \\
 24 & 52446.1233 &  0.0038 \\
 25 & 52446.2112 &  0.0037 \\
 47 & 52448.1379 & $-$0.0057 \\
\hline
 \multicolumn{3}{l}{$^a$ Cycle count since BJD 2452444.002.} \\
 \multicolumn{3}{l}{$^b$ $O-C$ calculated against equation
                    \ref{equ:kkreg1}.} \\
\end{tabular}
\end{center}
\end{table}

\begin{equation}
{\rm BJD (maximum)} = 2452444.0072(50) + 0.08801(18) E. \label{equ:kkreg1}
\end{equation}

   The times of the superhump maxima in this interval can be well
represented by the following quadratic equation:

\begin{eqnarray}
{\rm BJD} & {\rm (maximum)} = 2452444.0018 + 0.08880 E \nonumber \\
    & -1.69 \times 10^{-5} E^2. \label{equ:kkreg2}
\end{eqnarray}

   Nominal errors of the fit are not shown because of the small degree
of freedom in the data.  Although further detailed observations are
needed to exactly determine the period derivative, the presently measured
period change corresponds to
$\dot{P}$ = $-3.3 \times 10^{-5}$ d cycle$^{-1}$, or
$\dot{P}/P$ = $-3.7 \times 10^{-4}$.  Even if we allow most pessimistic
measurement errors of 0.01 d (14 min) for superhump maxima, the period
derivative varies only by $\sim$ 10\%.  We thereby adopted
$\dot{P}/P$ = $-3.7(4) \times 10^{-4}$ for KK Tel.

\subsection{Astrometry and Quiescent Counterpart}

   Astrometry of the outbursting KK Tel was performed on CCD
images taken by P. Nelson and T. Richards.  An average of measurements
of two images (UCAC1 system, 26 -- 70 reference stars; internal
dispersion of the measurements was $\sim$0$''$.3) has yielded
a position of 20$^h$ 28$^m$ 38$^s$.51, $-$52$^{\circ}$ 18$'$ 45$''$.0
(J2000.0, epoch=2002.461).  The position agrees with the USNO$-$A2.0 star at
20$^h$ 28$^m$ 38$^s$.471, $-$52$^{\circ}$ 18$'$ 45$''$.34
(epoch 1982.057 and magnitudes $r$ = 17.8, $b$ = 18.6), and the GSC$-$2.2
star at 20$^h$ 28$^m$ 38$^s$.456, $-$52$^{\circ}$ 18$'$ 45$''$.17
(epoch 1976.637 and magnitude $B_j$ = 19.24).  This identification
has confirmed a large outburst amplitude inferred by \citet{how91faintCV4}.

\subsection{KK Tel as an SU UMa-type Dwarf Nova}

\begin{figure*}
  \includegraphics[angle=0,width=16cm]{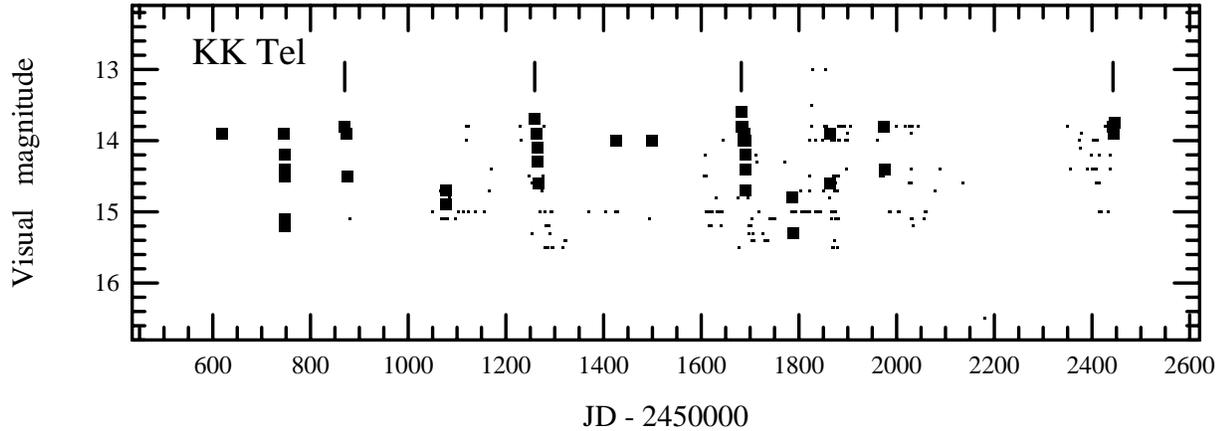}
  \caption{Long-term visual light curve of KK Tel.  Large and small dot
  represent positive and negative (upper limit) observations, respectively.
  Outbursts marked with vertical ticks are superoutbursts.}
  \label{fig:kklong}
\end{figure*}

   Figure \ref{fig:kklong} shows the long-term visual light curve of KK Tel.
Table \ref{tab:kkout} lists the observed outbursts.
Four well-defined superoutbursts (JD 2450870, 2451259, 2451682, 2452443)
with durations longer than 5 d are unambiguously identified (cf. section
\ref{sec:v877long}).  From the relatively regular intervals
between superoutbursts, we obtained a mean supercycle length of 394 d.
This supercycle length is relatively long among long-$P_{\rm SH}$ SU UMa-type
dwarf novae (e.g. \cite{nog97sxlmi}).\footnote{
  We generally refer to systems with $P_{\rm SH}$ longer than $\sim$0.08 d
  as long-$P_{\rm SH}$ SU UMa-type dwarf novae in this paper.  These periods
  are close to the lower edge of the CV period gap, in which the number
  density of the observed CVs is markedly decreased \citep{war95book},
  and the properties of the systems with these periods are considered to
  most strongly reflect their evolutionary stages among the SU UMa-type
  dwarf novae \citep{pod01amcvn}.
}
The long supercycle implies
a low mass-transfer rate \citep{ich94cycle,osa96review}, which is
consistent with the relatively large ($\sim$5.4 mag) outburst amplitude.

   The present secure superhump detection has confirmed the identification
by \citet{how91faintCV4} that the 0.084 d period is the orbital period.
By adopting $P_{\rm orb}$ = 0.084 d and the mean $P_{\rm SH}$ = 0.08808 d,
we obtain a fractional superhump excess
($\epsilon=P_{\rm SH}/P_{\rm orb}-1$) of $\sim$5\%.  Although the exact
determination of the orbital period would require additional
observations,\footnote{
  Our independent estimate from the extracted values from the figures
  in \citet{how91faintCV4} has confirmed that the suggested period is
  likely to be correct to $\sim$1\%.
}
the fractional superhump excess is not unusual for an SU UMa-type dwarf
nova with this $P_{\rm orb}$ \citep{mol92SHexcess,pat98evolution}.

   The appearance of strong hump features in quiescence suggests that
KK Tel would have a high orbital inclination.  This possibility is
further strengthened by the presence of broad Balmer emission lines
\citep{zwi95CVspec2}.  A close examination of the light curves obtained
during the present superoutburst, however, did not reveal the presence
of eclipses.  With a likely high inclination, KK Tel would be a promising
target to precisely determine physical parameters of a long-period
SU UMa-type dwarf nova.  There would also be a possibility of observing
a beat phenomenon between $P_{\rm SH}$ and $P_{\rm orb}$ during
superoutbursts (e.g. \cite{ish01rzleo}), although a short baseline
of the present observation did not allow us to confirm this phenomenon.

\begin{table}
\caption{List of Outbursts of KK Tel.}\label{tab:kkout}
\begin{center}
\begin{tabular}{ccccc}
\hline\hline
JD start$^a$ & JD end$^a$ & Max & Duration (d) & Type \\
\hline
50619.1 &   --    & 13.9 &  ?   & normal? \\
50746.9 & 50748.0 & 13.9 &  2   & normal \\
50870.2 & 50876.2 & 13.8 & $>$6 & super \\
51077.0 & 51077.1 & 14.7 &  1:  & normal \\
51259.4 & 51267.4 & 13.7 & $>$8 & super \\
51426.1 &   --    & 14.0 &  1:  & normal \\
51500.0 &   --    & 14.0 &  1:  & normal \\
51682.1 & 51691.4 & 13.6 & $>$9 & super \\
51787.2 & 51788.0 & 14.8 &  1:  & normal \\
51862.9 & 51864.0 & 13.9 &  2   & normal \\
51975.2 &   --    & 14.4 &  1:  & normal \\
52443.0 & 52248.2 & 13.8 & $>$5 & super \\
\hline
 \multicolumn{5}{l}{$^a$ JD$-$2450000.} \\
\end{tabular}
\end{center}
\end{table}

\section{PU CM\lowercase{a} = RX J0640$-$24}

   RX J0640-24 (=1RXS J064047.8-242305) is a ROSAT-selected cataclysmic
variable, normally at $B$ = 15.4.  Based on the detection at mag 11 on one
ESO B plate, the object has been suspected to be a dwarf nova (cf.
\cite{DownesCVatlas2}).
The object is located at 06$^h$ 40$^m$ 47$^s$.691,
$-$24$^{\circ}$ 23$'$ 14$''$.04 (GSC$-$2.2, J2000.0, epoch=1996.131,
$r$ = 14.87, $b$ = 16.17).
The object was monitored by one
of the authors (PS) using an AP-8 CCD camera attached to the 50-cm
reflector at the Iowa Robotic Observatory (IRO).  Schmeer (2000,
vsnet-alert 3956)\footnote{
http://www.kusastro.kyoto-u.ac.jp/vsnet/Mail/alert3000/\\msg00956.html.
} finally
discovered a new outburst on 2000 January 6.348 UT at unfiltered CCD magnitude
of 11.5, and established the dwarf nova classification.  Upon this alert,
we started time-resolved CCD photometry at Kyoto (TK and MU) and
Tsukuba (SK).  The Tsukuba observation further covered the later part
of the 2000 February outburst, which was detected by one of the authors
(R. Stubbings).

  The magnitudes of Kyoto CCD observations were determined
relative to GSC 6512.166 (Tycho-2 magnitude: $V=12.43\pm0.17$,
$B-V=+0.91\pm0.34$).  The magnitudes of Tsukuba observations were
determined relative to GSC 6512.908 (Tycho-2 magnitude: $V=9.10\pm0.01$,
$B-V=+1.31\pm0.03$).
The constancy of comparison star during the run was confirmed by
comparison with GSC 6508.1168 at both observatories.
The log of observations is given in Table \ref{tab:pucmalog}.
BM obtained another set of time-series photometry on 2002 May 27
(during the 2002 May long outburst).  Because of the different comparison
star system, average magnitude of BM's observation is not listed
in Table \ref{tab:pucmalog}.

\begin{table}
\begin{center}
\caption{Nightly averaged magnitudes of PU CMa}\label{tab:pucmalog}
\begin{tabular}{cccccc}
\hline\hline
BJD start$^a$ & BJD end$^a$ & Mean mag$^c$ & Error$^d$ & N$^d$ & Obs.$^e$\\
\hline
51551.202 & 51551.257 & 1.544 & 0.013 & 137 & K \\
51552.065 & 51552.206 & 5.406 & 0.025 &  63 & T \\
51552.123 & 51552.268 & 2.516 & 0.028 & 342 & K \\
51554.187 & 51554.260 & 3.010 & 0.092 &  60 & K \\
51558.190 & 51558.196 & 3.106 & 0.100 &  14 & K \\
51559.181 & 51559.199 & 3.597 & 0.057 &  49 & K \\
51561.156 & 51561.202 & 3.417 & 0.049 & 116 & K \\
51599.003 & 51599.018 & 4.331 & 0.012 &  10 & T \\
51599.963 & 51600.009 & 5.399 & 0.029 &  31 & T \\
51831.261 & 51831.264 & 2.929 & 0.262 &   8 & K \\
52422.178 & 52422.224 &  --   &  --   &  99 & M \\
\hline
  \multicolumn{6}{l}{$^a$ BJD$-$2400000.} \\
  \multicolumn{6}{l}{$^b$ Magnitude relative to GSC 6512.166 (Kyoto),} \\
  \multicolumn{6}{l}{\phantom{$^b$} GSC 6512.908 (Tsukuba).} \\
  \multicolumn{6}{l}{$^c$ Standard error of nightly average.} \\
  \multicolumn{6}{l}{$^d$ Number of frames.} \\
  \multicolumn{6}{l}{$^e$ K (Kyoto), T (Tsukuba), M (Monard).} \\
\end{tabular}
\end{center}
\end{table}

\begin{figure}
  \begin{center}
  \includegraphics[angle=0,width=7cm]{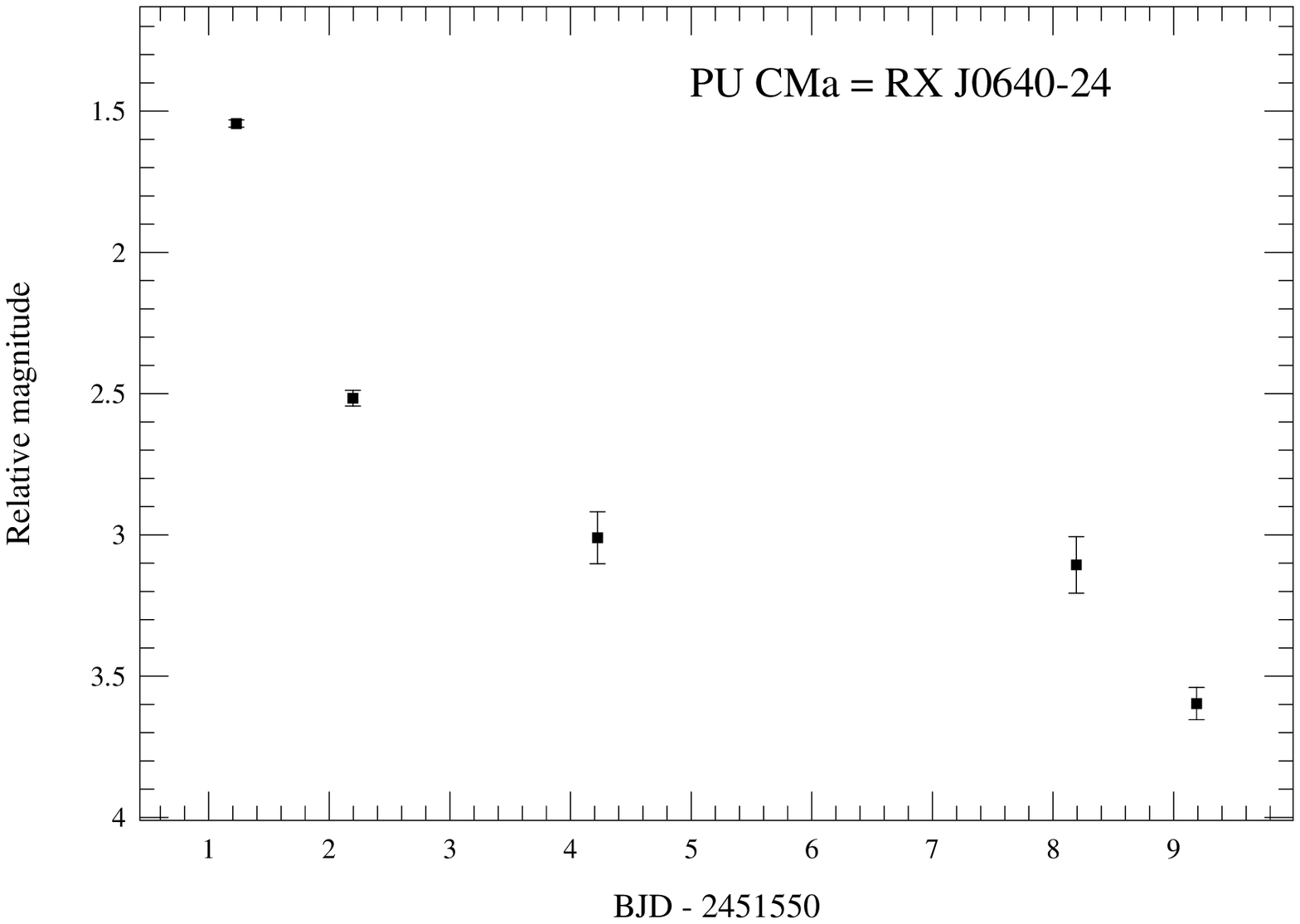} \\
  \includegraphics[angle=0,width=8cm]{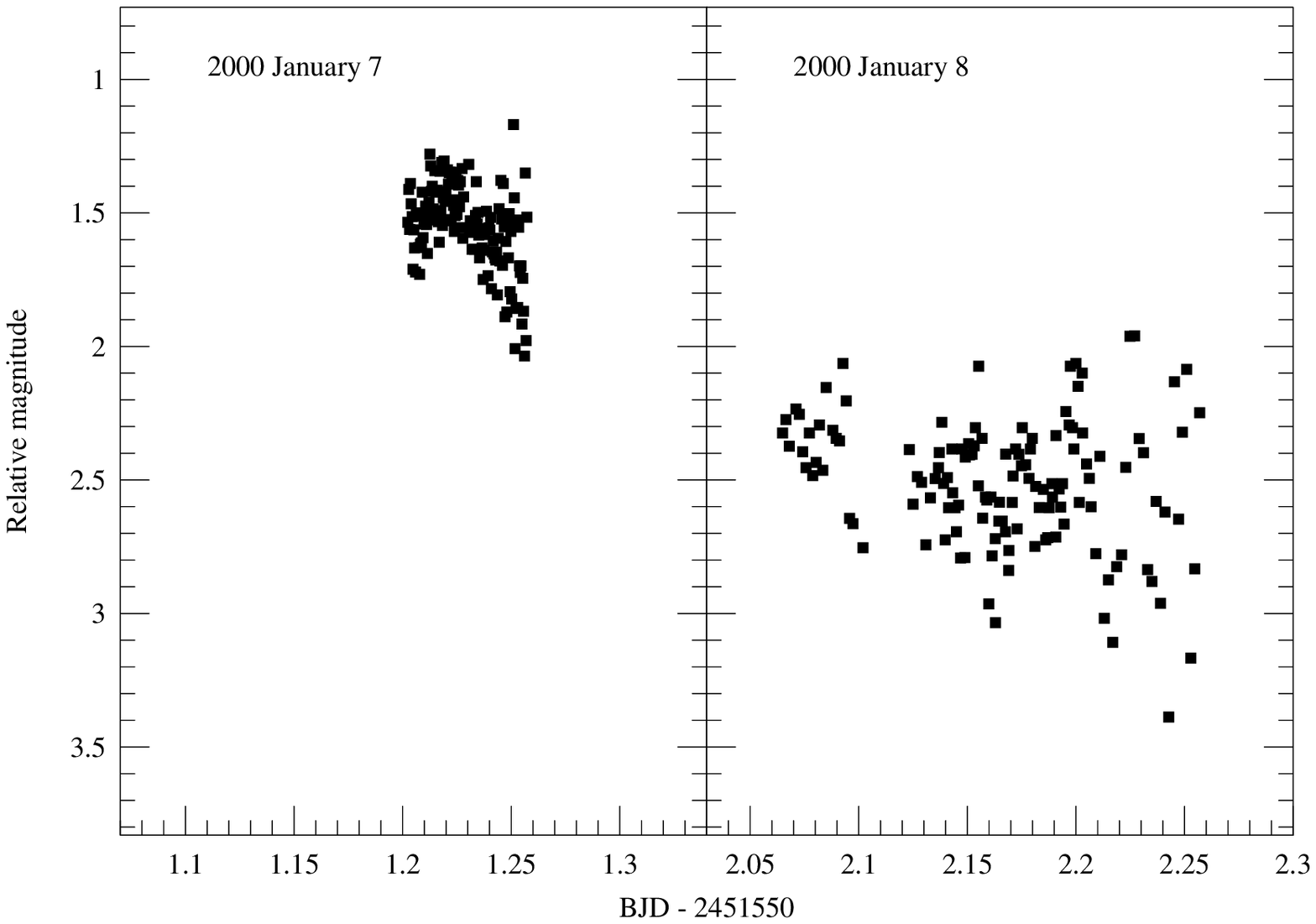}
  \end{center}
  \caption{(Upper) Light curve of the 2000 January outburst of PU CMa drawn
  from the Kyoto data.  Nightly averaged magnitudes are plotted with error
  bars representing standard errors.
  (Lower) Enlarged light curve of the 2000 January outburst.
  The Kyoto data on January 8 were binned to 0.002 d.}
  \label{fig:pufig1}
\end{figure}

   The upper panel of Figure \ref{fig:pufig1} shows the light curve of
the 2000 January outburst drawn from
the Kyoto data.  The light curve shows a rapid initial decline at a rate
of $\sim$1.0 mag d$^{-1}$.  The decline became slower as the object
approached its minimum magnitude.  The object returned to quiescence
within 4 d of the initial detection of the rise.  The object was observed
at quiescence (15.4 mag) two days before the outburst detection.
Such rapid evolution of the outburst suggests a normal outburst of an
SU UMa-type dwarf nova.  The lower panel of Figure \ref{fig:pufig1} shows
an enlarged light curve of the 2000 January outburst,
drawn from Kyoto and Tsukuba observations.  The Tsukuba observation on
January 8 were shifted by 2.966 mag in order to get the best match.
A smooth, slow modulation was observed on the first night (January 7),
while it became more irregular on January 8.  The time scale of the
January 7 variation looks slightly longer than the observing run (1.3 hr),
which may reflect some sort of a hump-like feature (e.g. orbital humps).

\begin{figure}
  \begin{center}
  \includegraphics[angle=0,width=8cm]{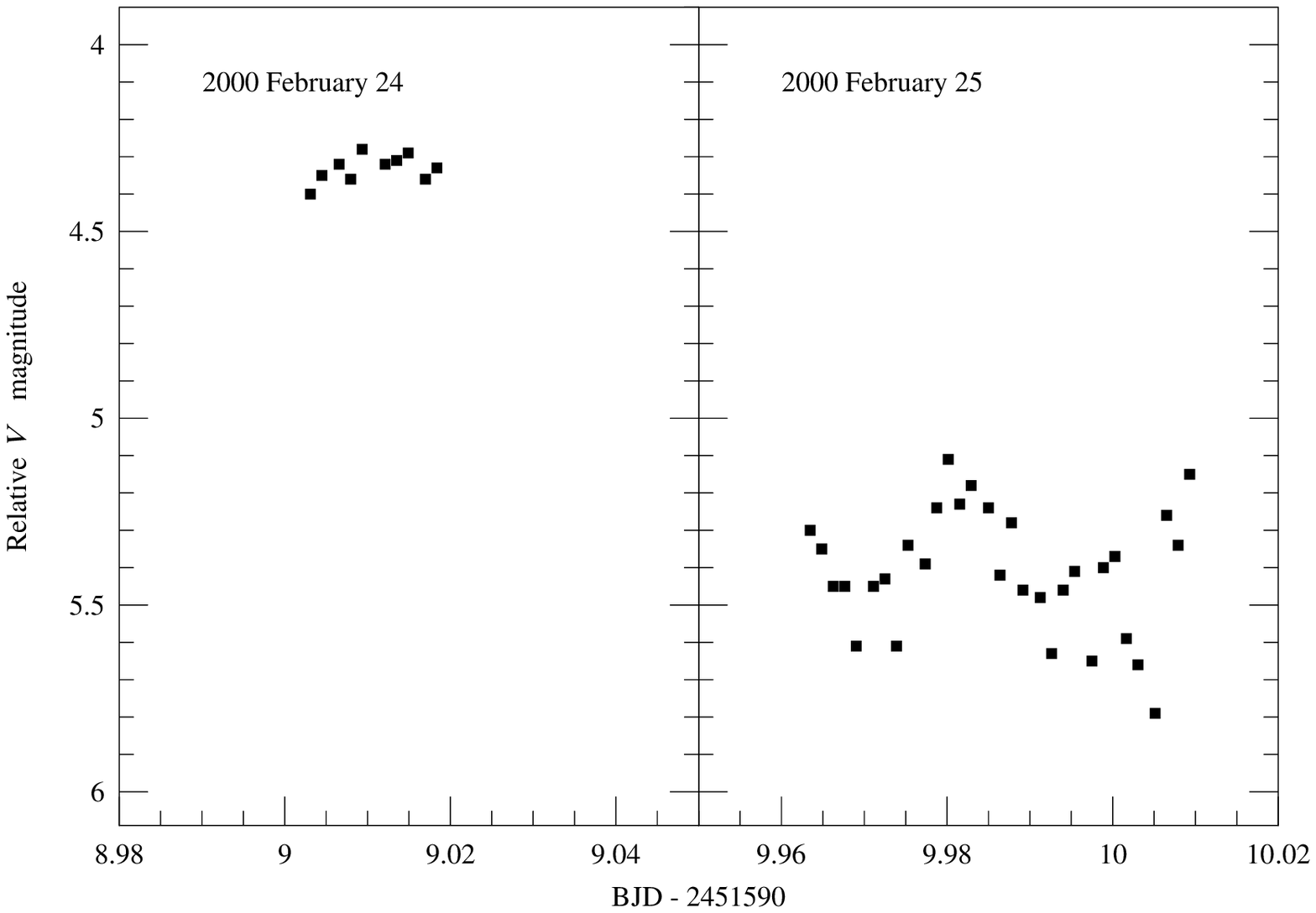} \\
  \includegraphics[angle=0,width=8cm]{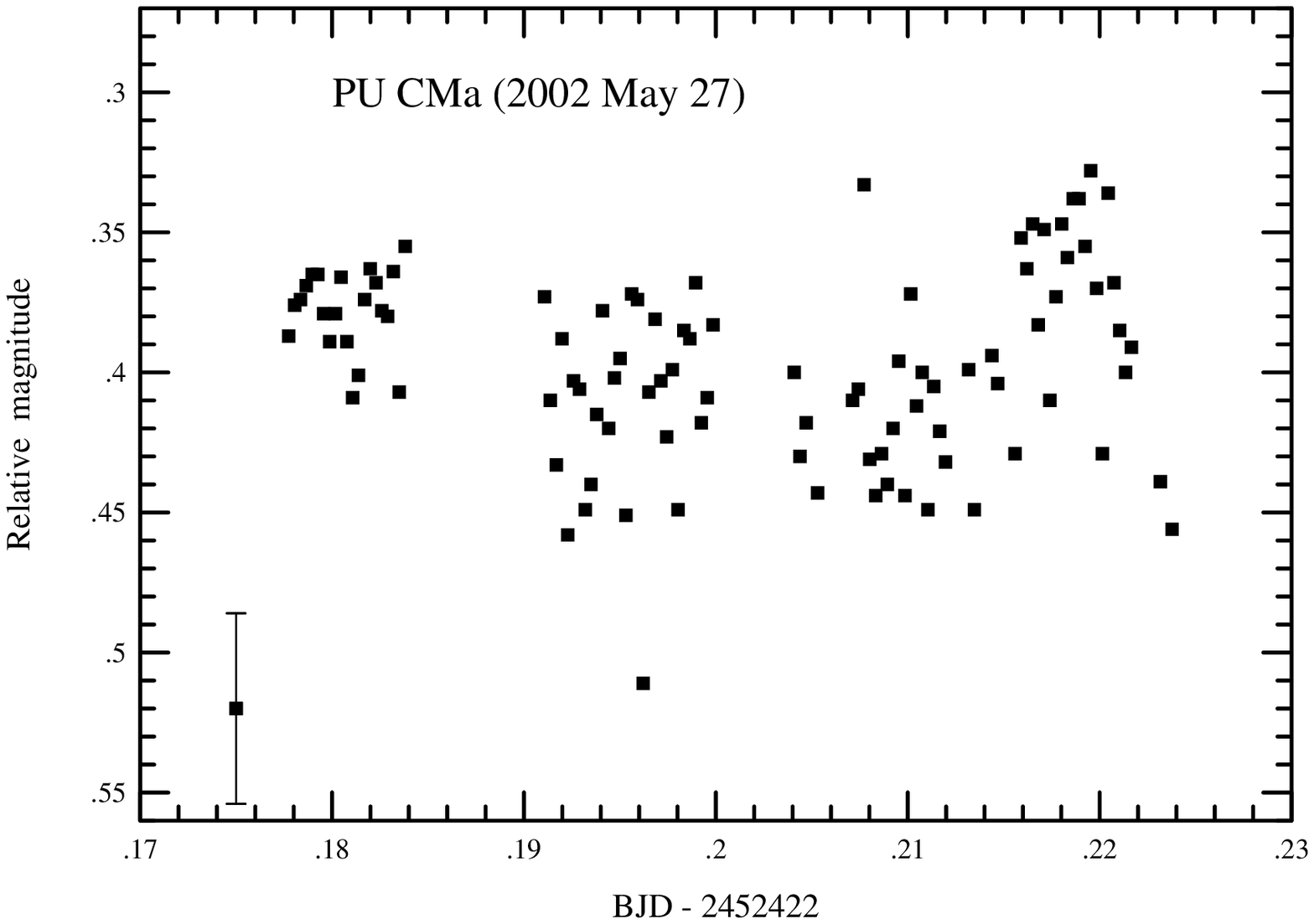} \\
  \end{center}
  \caption{(Upper) Enlarged light curve of the 2000 February outburst
  of PU CMa. (Lower) Time-resolved photometry of the 2002 May long
  outburst.}
  \label{fig:pufig3}
\end{figure}

   The upper panel of Figure \ref{fig:pufig3} shows an enlarged light curve
of the 2000 February outburst,
drawn Tsukuba observations.  Periodic-looking modulations were observed
on February 25, with a likely period of 0.032--0.033 d.  Although the
origin of the modulation is not clear, it may be related to some harmonic
of the orbital period.  The lower panel of Figure \ref{fig:pufig3} shows
a time-resolved light curve taken on 2002 May 27, the unique photometric
run during a long outburst.
Although there is a suggestion of $\sim$0.1 mag modulation, the unfavorable
photometric condition and the short seasonal visibility make it difficult
to draw a conclusion on the existence of superhumps.

\begin{figure*}
  \includegraphics[angle=0,width=16cm]{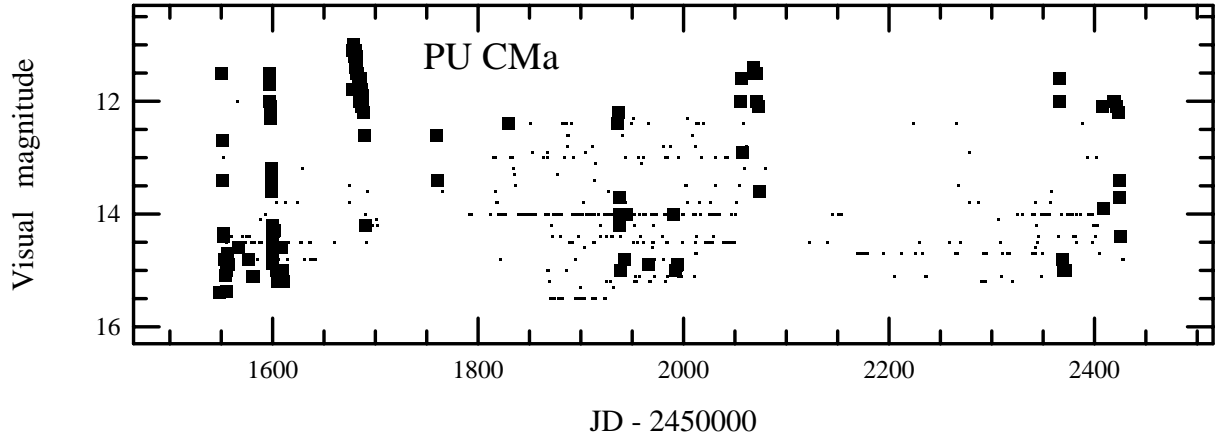}
  \caption{Long-term visual light curve of PU CMa.  Large and small dot
  represent positive and negative (upper limit) observations, respectively.
  Three long outbursts (JD 2451677, 2452056, 2452418) were detected.}
  \label{fig:pulong}
\end{figure*}

   Figure \ref{fig:pulong} depicts a long-term light curve of PU CMa,
based on visual observations (R. Stubbings, AP, BM, TW), as a part of
VSNET Collaboration.
All observers used $V$-band comparison stars.  In addition to short
outbursts as observed in 2000 January and February, three long outbursts
[JD 2451677 (2000 May), 2452056 (2001 June), 2452418 (2002 May)] were
detected.
The overall characteristics of outbursts are very reminiscent of an
SU UMa-type dwarf nova.  The interval of long, bright outburst (likely
superoutbursts) was 362--391 d, which may be the supercycle of this star.
If this periodicity continues, superoutbursts in the immediate future
will be difficult to observe.

   We also note that the second and third long outbursts were preceded
by a precursor 13 d and 11 d, respectively, before each outbursts.
Such appearance of precursors are relatively common in SU UMa-type
superoutbursts (e.g. \cite{mar79superoutburst,kat97tleo}),
and is naturally explained to
reflect two-step ignitions (thermal instability and tidal instability)
within the framework of the disk-instability model
\citep{osa89suuma,osa96review}.  If PU CMa would turn out to be an
SS Cyg-type dwarf nova rather than an SU UMa-type dwarf nova, such
a feature should require a different mechanism.

   Although definitive classification requires further time-resolved
photometry, we propose that PU CMa to be an excellent candidate for an
SU UMa-type dwarf nova.  This indication is also strengthened by the
recently reported orbital period of 0.05669(4) d
\citep{tho02kxaqlftcampucmav660herdmlyr}.  This period is one of the
shortest periods among the known SU UMa-type dwarf novae
\citep{tho02gwlibv844herdiuma}.  Most of the systems with similar
orbital periods show significant deviations (either related to WZ Sge-type
dwarf novae or ER UMa-type dwarf novae) from normal SU UMa-type
dwarf novae.  PU CMa, with its outburst properties strongly resembling
a normal SU UMa-type dwarf nova \citep{vog80suumastars,war85suuma},
may be the first object filling the gap between the extreme WZ Sge-type
and ER UMa-type systems.

\section{Superhump Period Changes in Long-Period Systems}

   We have shown that V877 Ara has a definitively large decrease of
the superhump period.  Even allowing for unusually large measurement errors
of superhump times, an even more striking decrease is inferred in KK Tel.
The periods of ``textbook" superhumps in usual
SU UMa-type dwarf novae have been known to decrease during a superoutburst
(e.g. \cite{war85suuma,pat93vyaqr}).  \citet{war85suuma,pat93vyaqr}
showed that the period derivative ($P_{\rm dot} = \dot{P}/P$) has
a rather common negative value ($\sim -5 \times 10^{-5}$).
This period change has been generally attributed to decreasing
apsidal motion due to a decreasing disk radius \citep{osa85SHexcess}
or inward propagation of the eccentricity wave \citep{lub92SH}.
\citet{kat01hvvir} systematically studied $P_{\rm dot}$ in SU UMa-type
dwarf novae.  Within the survey by \citet{kat01hvvir}, most of
long-$P_{\rm orb}$ SU UMa-type dwarf novae have been confirmed to show
a ``textbook" decrease of the
superhump periods.  On the other hand, short-period systems or infrequently
outbursting SU UMa-type systems have been found to predominantly
show an increase in the superhump periods.

\begin{figure*}
  \includegraphics[angle=0,width=15cm]{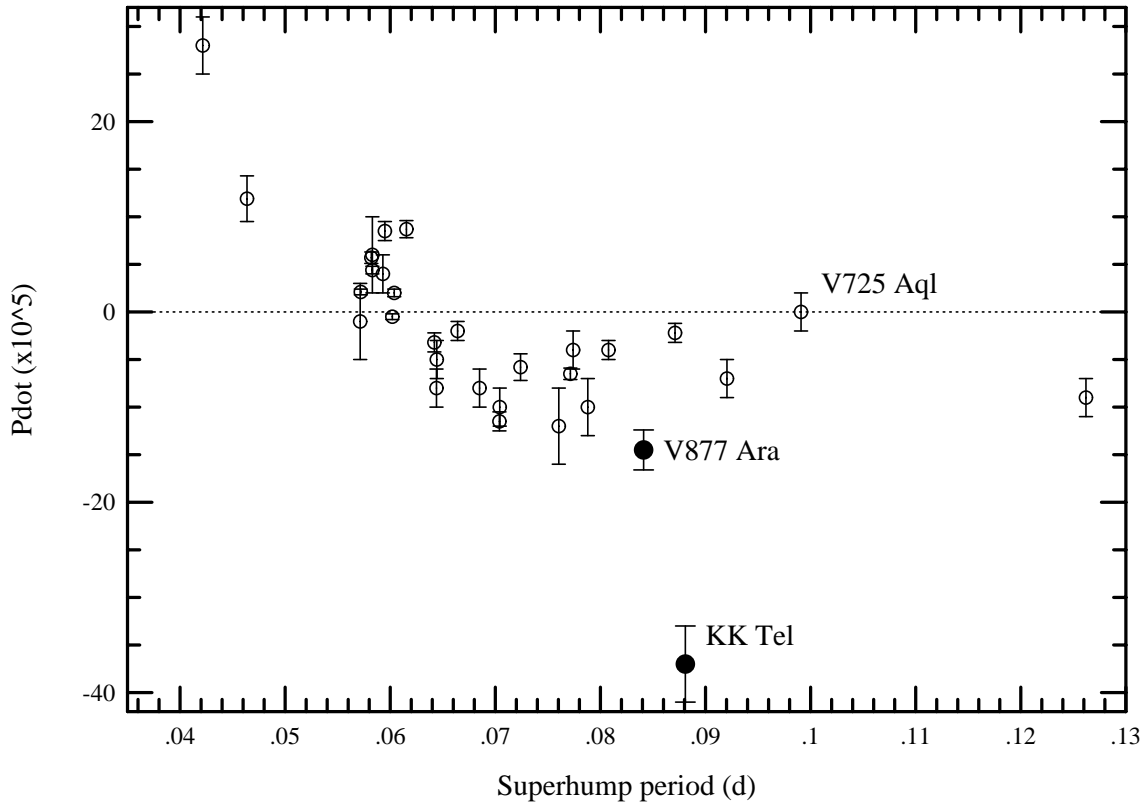}
  \caption{$\dot{P}/P$ versus $P_{\rm SH}$ for SU UMa-type dwarf novae.
  The open circles are from \citet{kat01hvvir}.  The data points of
  1RXS J232953.9+062814 and V725 Aql is from \citet{uem02j2329} and
  \citet{uem01v725aql}.
  }
  \label{fig:pdot}
\end{figure*}

   In recent years, some long-period SU UMa-type stars
(V725 Aql: \citet{uem01v725aql}; EF Peg: K. Matsumoto et al, in preparation)
have been known to show zero or marginally positive $P_{\rm dot}$.  While
this finding seems to give an impression that $P_{\rm dot}$ makes a minimum
around the period of 0.07--0.08 d, the discovery of a large negative
$P_{\rm dot}$ in long-period systems (V877 Ara and KK Tel) more implies a
previously neglected diversity of $P_{\rm dot}$ in long-period SU UMa-type
systems.  Figure \ref{fig:pdot} shows $\dot{P}/P$ versus $P_{\rm SH}$ for
SU UMa-type dwarf novae.  This diagram clearly shows the present of different
populations (i.e. systems with $P_{\rm dot} \sim$0 and systems with
definitely negative $P_{\rm dot}$) in long-period ($P_{\rm SH} >$ 0.08 d)
systems.

   The origin of different $\dot{P}/P$ among SU UMa-type dwarf novae was
first attributed to different tidal torques in different binary mass ratios
($q=M_2/M_1$)
[i.e. short-$P_{\rm SH}$ WZ Sge stars have the lowest $q$, which could
explain the small tidal effect on the accretion disk \citep{pat93vyaqr}].
\citet{kat98super} further proposed an idea that a large disk radius
in superoutbursting WZ Sge-type dwarf novae (cf. \cite{osa02wzsgehump})
enables outward propagation of eccentricity waves, which is responsible
for positive $\dot{P}/P$ in WZ Sge-type stars.  Recent discoveries
of nearly zero $\dot{P}/P$ in long-period, low mass-transfer rate
($\dot{M}$) systems (EF Peg and V725 Aql), however, seem to more favor
a new interpretation that $\dot{P}/P$ more reflects $\dot{M}$ (see also
\cite{kat01wxcet} for more discussion).  If the latter possibility is the
origin of different $\dot{P}/P$, the present discovery of a diversity of
$\dot{P}/P$ provides new evidence of a wide variety of $\dot{M}$ in
long-period SU UMa-type dwarf novae.  In short-period SU UMa-type
dwarf novae (and related systems), there is a well-established diversity
[WZ Sge-type stars and ER UMa stars, permanent superhumpers
\citep{dob92bklyn,ski93bklyn,rin96bklyn}].  There have been two recent
attempts to interpret this diversity by introducing an inner truncation
of the accretion disk and irradiation on the secondary star
\citep{bua01DNoutburst,bua02suumamodel} and by considering a decoupling
between the thermal and tidal instabilities in low $q$ systems
\citep{hel01eruma}.  Both models are expected to show smaller effects
in long-period SU UMa-type stars (e.g. $P_{\rm SH} >$ 0.08 d), and we
probably need a different explanation.  Further measurements of $\dot{P}/P$
in a larger SU UMa-type sample would become a key in solving this
still puzzling phenomenon.  There is a recently established and
addressed diversity in short-period SU UMa-type systems
\citep{war98CVreviewWyo,nog98CVevolution}, which has led to new insights
into fundamental problems of accretion disks
\citep{osa95eruma,osa95rzlmi,bua01DNoutburst,bua02suumamodel,hel01eruma}.
The present discovery of a diversity in long-period SU UMa-type systems
would become an additional step toward full understanding the dwarf nova
and CV phenomenon.

\section*{Acknowledgments}

This work is partly supported by a grant-in aid [13640239 (TK),
14740131 (HY)] from the Japanese Ministry of Education, Culture, Sports,
Science and Technology.
Part of this work is supported by a Research Fellowship of the
Japan Society for the Promotion of Science for Young Scientists (MU).
The CCD operation of the Bronberg Observatory is partly sponsored by
the Center for Backyard Astrophysics.
The CCD operation by Peter Nelson is on loan from the AAVSO,
funded by the Curry Foundation.
P. Schmeer's observations were made with the Iowa Robotic Observatory,
and he wishes to thank Robert Mutel and his students.
This research has made use of the Digitized Sky Survey producted by STScI, 
the ESO Skycat tool, the VizieR catalogue access tool.

\label{lastpage}


\begin{thebibliography}{}

\bibitem[\protect\citeauthoryear{Bailey \& Ward}{Bailey \&
  Ward}{1981}]{bai81oycar}
Bailey J.,  Ward M.,  1981, MNRAS, 194, 17

\bibitem[\protect\citeauthoryear{Buat-M\'{e}nard \& Hameury}{Buat-M\'{e}nard \&
  Hameury}{2002}]{bua02suumamodel}
Buat-M\'{e}nard V.,  Hameury J.-M.,  2002, A\&A, 386, 891

\bibitem[\protect\citeauthoryear{Buat-M\'{e}nard, Hameury \&
  Lasota}{Buat-M\'{e}nard et~al.}{2001}]{bua01DNoutburst}
Buat-M\'{e}nard V.,  Hameury J.-M.,    Lasota J.-P.,  2001, A\&A, 366, 612

\bibitem[\protect\citeauthoryear{Cook \& Warner}{Cook \&
  Warner}{1981}]{coo81zchapdot}
Cook M.~C.,  Warner B.,  1981, MNRAS, 196, 55P

\bibitem[\protect\citeauthoryear{Cook \& Warner}{Cook \&
  Warner}{1984}]{coo84zcha}
Cook M.~C.,  Warner B.,  1984, MNRAS, 207, 705

\bibitem[\protect\citeauthoryear{Dobrzycka \& Howell}{Dobrzycka \&
  Howell}{1992}]{dob92bklyn}
Dobrzycka D.,  Howell S.~B.,  1992, ApJ, 388, 614

\bibitem[\protect\citeauthoryear{Downes, Webbink \& Shara}{Downes
  et~al.}{1997}]{DownesCVatlas2}
Downes R.,  Webbink R.~F.,    Shara M.~M.,  1997, PASP, 109, 345

\bibitem[\protect\citeauthoryear{Hellier}{Hellier}{2001a}]{hel01book}
Hellier C.,  2001a, Cataclysmic Variable Stars: how and why they vary
(Berlin: Springer-Verlag)

\bibitem[\protect\citeauthoryear{Hellier}{Hellier}{2001b}]{hel01eruma}
Hellier C.,  2001b, PASP, 113, 469

\bibitem[\protect\citeauthoryear{Hirose \& Osaki}{Hirose \&
  Osaki}{1990}]{hir90SHexcess}
Hirose M.,  Osaki Y.,  1990, PASJ, 42, 135

\bibitem[\protect\citeauthoryear{Howell \& Szkody}{Howell \&
  Szkody}{1988}]{how88faintCV1}
Howell S.,  Szkody P.,  1988, PASP, 100, 224

\bibitem[\protect\citeauthoryear{Howell \& Blanton}{Howell \&
  Blanton}{1993}]{how93arcncaypscdvuma}
Howell S.~B.,  Blanton S.~A.,  1993, AJ, 106, 311

\bibitem[\protect\citeauthoryear{Howell, DeYoung, Mattei, Foster, Szkody,
  Cannizzo, Walker \& Fierce}{Howell et~al.}{1996}]{how96alcom}
Howell S.~B.,  DeYoung J.~A.,  Mattei J.~A.,  Foster G.,  Szkody P.,  Cannizzo
  J.~K.,  Walker G.,    Fierce E.,  1996, AJ, 111, 2367

\bibitem[\protect\citeauthoryear{Howell, Dobrzycka, Szkody \& Kreidl}{Howell
  et~al.}{1991}]{how91faintCV4}
Howell S.~B.,  Dobrzycka D.,  Szkody P.,    Kreidl T.~J.,  1991, PASP, 103,
  300

\bibitem[\protect\citeauthoryear{Howell, Mitchell \& Warnock}{Howell
  et~al.}{1987}]{how87dvuma}
Howell S.~B.,  Mitchell K.~J.,    Warnock A.~I.,  1987, PASP, 99, 126

\bibitem[\protect\citeauthoryear{Howell, Reyes, Ashley, Harrop-Allin \&
  Warner}{Howell et~al.}{1996}]{how96tvcrv}
Howell S.~B.,  Reyes A.~L.,  Ashley R.,  Harrop-Allin M.~K.,    Warner B.,
  1996, MNRAS, 282, 623

\bibitem[\protect\citeauthoryear{Howell \& Szkody}{Howell \&
  Szkody}{1990}]{how90highgalCV}
Howell S.~B.,  Szkody P.,  1990, ApJ, 356, 623

\bibitem[\protect\citeauthoryear{Howell, Szkody \& Cannizzo}{Howell
  et~al.}{1995}]{how95TOAD}
Howell S.~B.,  Szkody P.,    Cannizzo J.~K.,  1995, ApJ, 439, 337

\bibitem[\protect\citeauthoryear{Howell, Szkody, Kreidl, Mason \&
  Puchnarewicz}{Howell et~al.}{1990}]{how90faintCV3}
Howell S.~B.,  Szkody P.,  Kreidl T.~J.,  Mason K.~O.,    Puchnarewicz E.~M.,
  1990, PASP, 102, 758

\bibitem[\protect\citeauthoryear{Howell, Warnock, Mason, Reichert \&
  Kreidl}{Howell et~al.}{1988}]{how88dvuma}
Howell S.~B.,  Warnock A.,  Mason K.~O.,  Reichert G.~A.,    Kreidl T.~J.,
  1988, MNRAS, 233, 79

\bibitem[\protect\citeauthoryear{Ichikawa \& Osaki}{Ichikawa \&
  Osaki}{1994}]{ich94cycle}
Ichikawa S.,  Osaki Y.,  1994, in Duschl W.~J.,  Frank J.,  Meyer F.,
  Meyer-Hofmeister E.,   Tscharnuter W.~M.,  eds, Theory of Accretion Disks-2
  (Dordrecht: Kluwer Academic Publishers), p.~169

\bibitem[\protect\citeauthoryear{Ishioka, Kato, Uemura, Iwamatsu, Matsumoto,
  Stubbings, Mennickent, Billings, Kiyota, Masi, Pietz, Nov\'{a}k, Martin,
  Oksanen, Moilanen, Torii, Kinugasa \& Kawakita}{Ishioka
  et~al.}{2001}]{ish01rzleo}
Ishioka R.,  Kato T.,  Uemura M.,  Iwamatsu H.,  Matsumoto K.,  Stubbings R.,
  Mennickent R.,  Billings G.~W.,  Kiyota S.,  Masi G.,  Pietz J.,  Nov\'{a}k
  R.,  Martin B.,  Oksanen A.,  Moilanen M.,  Torii K.,  Kinugasa K.,
  Kawakita H.,  2001, PASJ, 53, 905

\bibitem[\protect\citeauthoryear{Kato}{Kato}{1997}]{kat97tleo}
Kato T.,  1997, PASJ, 49, 583

\bibitem[\protect\citeauthoryear{Kato \& Kunjaya}{Kato \&
  Kunjaya}{1995}]{kat95eruma}
Kato T.,  Kunjaya C.,  1995, PASJ, 47, 163

\bibitem[\protect\citeauthoryear{Kato, Matsumoto, Nogami, Morikawa \&
  Kiyota}{Kato et~al.}{2001}]{kat01wxcet}
Kato T.,  Matsumoto K.,  Nogami D.,  Morikawa K.,    Kiyota S.,  2001, PASJ,
  53, 893

\bibitem[\protect\citeauthoryear{Kato, Nogami \& Baba}{Kato
  et~al.}{1996}]{kat96diuma}
Kato T.,  Nogami D.,    Baba H.,  1996, PASJ, 48, L93

\bibitem[\protect\citeauthoryear{Kato, Nogami, Baba \& Matsumoto}{Kato
  et~al.}{1998}]{kat98super}
Kato T.,  Nogami D.,  Baba H.,    Matsumoto K.,  1998, in Howell S.,  Kuulkers
  E.,   Woodward C.,  eds, ASP Conf. Ser. 137, Wild Stars in the Old West
  (San Francisco: ASP), p.~9

\bibitem[\protect\citeauthoryear{Kato, Nogami, Baba, Matsumoto, Arimoto, Tanabe
  \& Ishikawa}{Kato et~al.}{1996}]{kat96alcom}
Kato T.,  Nogami D.,  Baba H.,  Matsumoto K.,  Arimoto J.,  Tanabe K.,
  Ishikawa K.,  1996, PASJ, 48, L21

\bibitem[\protect\citeauthoryear{Kato, Nogami, Matsumoto \& Baba}{Kato
  et~al.}{1997}]{kat97egcnc}
Kato T.,  Nogami D.,  Matsumoto K.,    Baba H.,  1997, preprint,
  ftp://vsnet.kusastro.kyoto-u.ac.jp/pub/vsnet/\\ preprints/EG\_Cnc/

\bibitem[\protect\citeauthoryear{Kato, Sekine \& Hirata}{Kato
  et~al.}{2001}]{kat01hvvir}
Kato T.,  Sekine Y.,    Hirata R.,  2001, PASJ, 53, 1191

\bibitem[\protect\citeauthoryear{Lopez}{Lopez}{1985}]{lop85CVastrometry}
Lopez C.~E.,  1985, Inf. Bull. Var. Stars, 2837

\bibitem[\protect\citeauthoryear{Lubow}{Lubow}{1991}]{lub91SHa}
Lubow S.~H.,  1991, ApJ, 381, 259

\bibitem[\protect\citeauthoryear{Lubow}{Lubow}{1992}]{lub92SH}
Lubow S.~H.,  1992, ApJ, 401, 317

\bibitem[\protect\citeauthoryear{Marino \& Walker}{Marino \&
  Walker}{1979}]{mar79superoutburst}
Marino B.~F.,  Walker W. S.~G.,  1979, in Bateson F.~M.,  Smak J.,   Urch
  J.~H.,  eds, IAU Colloq. 46, Changing Trends in Variable Star Research
  (Univ. of Waikato, Hamilton, N. Z.), p.~29

\bibitem[\protect\citeauthoryear{Matsumoto, Nogami, Kato \& Baba}{Matsumoto
  et~al.}{1998}]{mat98egcnc}
Matsumoto K.,  Nogami D.,  Kato T.,    Baba H.,  1998, PASJ, 50, 405

\bibitem[\protect\citeauthoryear{Molnar \& Kobulnicky}{Molnar \&
  Kobulnicky}{1992}]{mol92SHexcess}
Molnar L.~A.,  Kobulnicky H.~A.,  1992, ApJ, 392, 678

\bibitem[\protect\citeauthoryear{Mukai, Mason, Howell, Allington-Smith,
  Callnan, Charles, Hassall, Machin, Naylor, Smale \& van Paradijs}{Mukai
  et~al.}{1990}]{muk90faintCV}
Mukai K.,  Mason K.~O.,  Howell S.~B.,  Allington-Smith J.,  Callnan P.~J.,
  Charles P.~A.,  Hassall B. J.~M.,  Machin G.,  Naylor T.,  Smale A.~P.,
  van Paradijs J.,  1990, MNRAS, 245, 385

\bibitem[\protect\citeauthoryear{Murray}{Murray}{1996}]{mur96SPHtidal}
Murray J.~R.,  1996, MNRAS, 279, 402

\bibitem[\protect\citeauthoryear{Murray}{Murray}{1998}]{mur98SH}
Murray J.~R.,  1998, MNRAS, 297, 323

\bibitem[\protect\citeauthoryear{Murray}{Murray}{2000}]{mur00SHprecession}
Murray J.~R.,  2000, MNRAS, 314, 1P

\bibitem[\protect\citeauthoryear{Nogami}{Nogami}{1998}]{nog98CVevolution}
Nogami D.,  1998, in Howell S.,  Kuulkers E.,   Woodward C.,  eds,
  ASP Conf. Ser. 137, Wild Stars in the Old West (San Francisco: ASP), p.~495

\bibitem[\protect\citeauthoryear{Nogami, Kato, Baba, Matsumoto, Arimoto, Tanabe
  \& Ishikawa}{Nogami et~al.}{1997}]{nog97alcom}
Nogami D.,  Kato T.,  Baba H.,  Matsumoto K.,  Arimoto J.,  Tanabe K.,
  Ishikawa K.,  1997, ApJ, 490, 840

\bibitem[\protect\citeauthoryear{Nogami, Kato, Masuda \& Hirata}{Nogami
  et~al.}{1995a}]{nog95v1159ori}
Nogami D.,  Kato T.,  Masuda S.,    Hirata R.,  1995a, Inf. Bull. Var. Stars,
  4155

\bibitem[\protect\citeauthoryear{Nogami, Kato, Masuda, Hirata, Matsumoto,
  Tanabe \& Yokoo}{Nogami et~al.}{1995b}]{nog95rzlmi}
Nogami D.,  Kato T.,  Masuda S.,  Hirata R.,  Matsumoto K.,  Tanabe K.,
  Yokoo T.,  1995b, PASJ, 47, 897

\bibitem[\protect\citeauthoryear{Nogami, Masuda \& Kato}{Nogami
  et~al.}{1997}]{nog97sxlmi}
Nogami D.,  Masuda S.,    Kato T.,  1997, PASP, 109, 1114

\bibitem[\protect\citeauthoryear{O'Donoghue}{O'Donoghue}{1986}]{odo86zchadiskr%
adius}
O'Donoghue D.,  1986, MNRAS, 220, 23

\bibitem[\protect\citeauthoryear{Osaki}{Osaki}{1985}]{osa85SHexcess}
Osaki Y.,  1985, A\&A, 144, 369

\bibitem[\protect\citeauthoryear{Osaki}{Osaki}{1989}]{osa89suuma}
Osaki Y.,  1989, PASJ, 41, 1005

\bibitem[\protect\citeauthoryear{Osaki}{Osaki}{1995a}]{osa95eruma}
Osaki Y.,  1995a, PASJ, 47, L11

\bibitem[\protect\citeauthoryear{Osaki}{Osaki}{1995b}]{osa95rzlmi}
Osaki Y.,  1995b, PASJ, 47, L25

\bibitem[\protect\citeauthoryear{Osaki}{Osaki}{1996}]{osa96review}
Osaki Y.,  1996, PASP, 108, 39

\bibitem[\protect\citeauthoryear{Osaki \& Meyer}{Osaki \&
  Meyer}{2002}]{osa02wzsgehump}
Osaki Y.,  Meyer F.,  2002, A\&A, 383, 574

\bibitem[\protect\citeauthoryear{Patterson}{Patterson}{1998}]{pat98evolution}
Patterson J.,  1998, PASP, 110, 1132

\bibitem[\protect\citeauthoryear{Patterson, Augusteijn, Harvey, Skillman,
  Abbott \& Thorstensen}{Patterson et~al.}{1996}]{pat96alcom}
Patterson J.,  Augusteijn T.,  Harvey D.~A.,  Skillman D.~R.,  Abbott T. M.~C.,
     Thorstensen J.,  1996, PASP, 108, 748

\bibitem[\protect\citeauthoryear{Patterson, Bond, Grauer, Shafter \&
  Mattei}{Patterson et~al.}{1993}]{pat93vyaqr}
Patterson J.,  Bond H.~E.,  Grauer A.~D.,  Shafter A.~W.,    Mattei J.~A.,
  1993, PASP, 105, 69

\bibitem[\protect\citeauthoryear{Patterson, Jablonski, Koen, O'Donoghue \&
  Skillman}{Patterson et~al.}{1995}]{pat95v1159ori}
Patterson J.,  Jablonski F.,  Koen C.,  O'Donoghue D.,    Skillman D.~R.,
  1995, PASP, 107, 1183

\bibitem[\protect\citeauthoryear{Patterson, Kemp, Skillman, Harvey, Shafter,
  Vanmunster, Jensen, Fried, Kiyota, Thorstensen \& Taylor}{Patterson
  et~al.}{1998}]{pat98egcnc}
Patterson J.,  Kemp J.,  Skillman D.~R.,  Harvey D.~A.,  Shafter A.~W.,
  Vanmunster T.,  Jensen L.,  Fried R.,  Kiyota S.,  Thorstensen J.~R.,
  Taylor C.~J.,  1998, PASP, 110, 1290

\bibitem[\protect\citeauthoryear{Podsiadlowski, Han \& Rappaport}{Podsiadlowski
  et~al.}{2001}]{pod01amcvn}
Podsiadlowski P.,  Han Z.,    Rappaport S.,  2001, MNRAS,
  submitted (astro-ph/0109171)

\bibitem[\protect\citeauthoryear{Ringwald, Thorstensen, Honeycutt \&
  Robertson}{Ringwald et~al.}{1996}]{rin96bklyn}
Ringwald F.~A.,  Thorstensen J.~R.,  Honeycutt R.~K.,    Robertson J.~W.,
  1996, MNRAS, 278, 125

\bibitem[\protect\citeauthoryear{Robertson, Honeycutt \& Turner}{Robertson
  et~al.}{1995}]{rob95eruma}
Robertson J.~W.,  Honeycutt R.~K.,    Turner G.~W.,  1995, PASP, 107, 443

\bibitem[\protect\citeauthoryear{Skillman, Krajci, Beshore, Patterson, Kemp,
  Starkey, Oksanen, Vanmunster, Martin \& Rea}{Skillman
  et~al.}{2002}]{ski02j2329}
Skillman D.~R.,  Krajci T.,  Beshore E.,  Patterson J.,  Kemp J.,  Starkey D.,
  Oksanen A.,  Vanmunster T.,  Martin B.,    Rea R.,  2002, PASP, 114, 630

\bibitem[\protect\citeauthoryear{Skillman \& Patterson}{Skillman \&
  Patterson}{1993}]{ski93bklyn}
Skillman D.~R.,  Patterson J.,  1993, ApJ, 417, 298

\bibitem[\protect\citeauthoryear{Smak}{Smak}{1979}]{sma79zcha}
Smak J.,  1979, Acta Astron., 29, 309

\bibitem[\protect\citeauthoryear{Smak}{Smak}{1985}]{sma85SHzcha}
Smak J.,  1985, Acta Astron., 35, 1

\bibitem[\protect\citeauthoryear{Stellingwerf}{Stellingwerf}{1978}]{PDM}
Stellingwerf R.~F.,  1978, ApJ, 224, 953

\bibitem[\protect\citeauthoryear{Szkody, Howell, Mateo \& Kreidl}{Szkody
  et~al.}{1989}]{szk89faintCV2}
Szkody P.,  Howell S.~B.,  Mateo M.,    Kreidl T.~J.,  1989, PASP, 101, 899

\bibitem[\protect\citeauthoryear{Thorstensen \& Fenton}{Thorstensen \&
  Fenton}{2002}]{tho02kxaqlftcampucmav660herdmlyr}
Thorstensen J.~R.,  Fenton W.~H.,  2002, PASP, in press, (astro-ph/0209172)

\bibitem[\protect\citeauthoryear{Thorstensen, Patterson, Kemp \&
  Vennes}{Thorstensen et~al.}{2002}]{tho02gwlibv844herdiuma}
Thorstensen J.~R.,  Patterson J.~O.,  Kemp J.,    Vennes S.,  2002, PASP,
  114, 1108

\bibitem[\protect\citeauthoryear{Truss, Murray \& Wynn}{Truss
  et~al.}{2001}]{tru01DNsuperoutburst}
Truss M.~R.,  Murray J.~R.,    Wynn G.~A.,  2001, MNRAS, 324, 1P

\bibitem[\protect\citeauthoryear{Truss, Murray, Wynn \& Edgar}{Truss
  et~al.}{2000}]{tru00DNoutburst}
Truss M.~R.,  Murray J.~R.,  Wynn G.~A.,    Edgar R.~G.,  2000, MNRAS, 319,
  467

\bibitem[\protect\citeauthoryear{Uemura, Kato, Ishioka, Yamaoka, Schmeer,
  Starkey, Torii, Kawai, Urata, Kohama, Yoshida, Ayani, Kawabata, Tanabe,
  Matsumoto, Kiyota, Pietz, Vanmunster, Krajci, Oksanen \& Giambersio}{Uemura
  et~al.}{2002a}]{uem02j2329letter}
Uemura M.,  Kato T.,  Ishioka R.,  Yamaoka H.,  Schmeer P.,  Starkey D.~R.,
  Torii K.,  Kawai N.,  Urata Y.,  Kohama M.,  Yoshida A.,  Ayani K.,  Kawabata
  T.,  Tanabe K.,  Matsumoto K.,  Kiyota S.,  Pietz J.,  Vanmunster T.,  Krajci
  T.,  Oksanen A.,    Giambersio A.,  2002a, PASJ, 54, L15

\bibitem[\protect\citeauthoryear{Uemura, Kato, Ishioka, Yamaoka, Schmeer,
  Starkey, Torii, Kawai, Urata, Kohama, Yoshida, Ayani, Kawabata, Tanabe,
  Matsumoto, Kiyota, Pietz, Vanmunster, Krajci, Oksanen \& Giambersio}{Uemura
  et~al.}{2002b}]{uem02j2329}
Uemura M.,  Kato T.,  Ishioka R.,  Yamaoka H.,  Schmeer P.,  Starkey D.~R.,
  Torii K.,  Kawai N.,  Urata Y.,  Kohama M.,  Yoshida A.,  Ayani K.,  Kawabata
  T.,  Tanabe K.,  Matsumoto K.,  Kiyota S.,  Pietz J.,  Vanmunster T.,  Krajci
  T.,  Oksanen A.,    Giambersio A.,  2002b, PASJ, 54, 599

\bibitem[\protect\citeauthoryear{Uemura, Kato, Pavlenko, Baklanov \&
  Pietz}{Uemura et~al.}{2001}]{uem01v725aql}
Uemura M.,  Kato T.,  Pavlenko E.,  Baklanov A.,    Pietz J.,  2001, PASJ, 53,
  539

\bibitem[\protect\citeauthoryear{Vogt}{Vogt}{1974}]{vog74vwhyi}
Vogt N.,  1974, A\&A, 36, 369

\bibitem[\protect\citeauthoryear{Vogt}{Vogt}{1980}]{vog80suumastars}
Vogt N.,  1980, A\&A, 88, 66

\bibitem[\protect\citeauthoryear{Vogt}{Vogt}{1982}]{vog82zcha}
Vogt N.,  1982, ApJ, 252, 653

\bibitem[\protect\citeauthoryear{Vogt}{Vogt}{1983a}]{vog83oycar}
Vogt N.,  1983a, A\&A, 128, 29

\bibitem[\protect\citeauthoryear{Vogt}{Vogt}{1983b}]{vog83lateSH}
Vogt N.,  1983b, A\&A, 118, 95

\bibitem[\protect\citeauthoryear{Vogt \& Bateson}{Vogt \&
  Bateson}{1982}]{vog82atlas}
Vogt N.,  Bateson F.~M.,  1982, A\&AS, 48, 383

\bibitem[\protect\citeauthoryear{Vogt, Schoembs, Krzeminski \& Pedersen}{Vogt
  et~al.}{1981}]{vog81oycar}
Vogt N.,  Schoembs R.,  Krzeminski W.,    Pedersen H.,  1981, A\&A, 94, L29

\bibitem[\protect\citeauthoryear{Warner}{Warner}{1974}]{war74zcha}
Warner B.,  1974, MNRAS, 168, 235

\bibitem[\protect\citeauthoryear{Warner}{Warner}{1975}]{war75v436cen}
Warner B.,  1975, MNRAS, 173, 37

\bibitem[\protect\citeauthoryear{Warner}{Warner}{1985}]{war85suuma}
Warner B.,  1985, in Eggelton P.~P.,  Pringle J.~E.,  eds, Interacting Binaries
  (Dordrecht: D. Reidel Publishing Company), p.~367

\bibitem[\protect\citeauthoryear{Warner}{Warner}{1995a}]{war95book}
Warner B.,  1995a, Cataclysmic Variable Stars
(Cambridge: Cambridge University Press)

\bibitem[\protect\citeauthoryear{Warner}{Warner}{1995b}]{war95suuma}
Warner B.,  1995b, Ap\&SS, 226, 187

\bibitem[\protect\citeauthoryear{Warner}{Warner}{1998}]{war98CVreviewWyo}
Warner B.,  1998, in Howell S.,  Kuulkers E.,   Woodward C.,  eds,
  ASP Conf. Ser. 137, Wild Stars in the Old West (San Francisco: ASP), p.~2

\bibitem[\protect\citeauthoryear{Warner \& Brickhill}{Warner \&
  Brickhill}{1974}]{war74vwhyi}
Warner B.,  Brickhill A.~J.,  1974, MNRAS, 166, 673

\bibitem[\protect\citeauthoryear{Whitehurst}{Whitehurst}{1988}]{whi88tidal}
Whitehurst R.,  1988, MNRAS, 232, 35

\bibitem[\protect\citeauthoryear{Woudt \& Warner}{Woudt \&
  Warner}{2001}]{wou01v359cenxzeriyytel}
Woudt P.~A.,  Warner B.,  2001, MNRAS, 328, 159

\bibitem[\protect\citeauthoryear{Zwitter \& Munari}{Zwitter \&
  Munari}{1995}]{zwi95CVspec2}
Zwitter T.,  Munari U.,  1995, A\&AS, 114, 575

\end{thebibliography}
\end{document}